# Using parametric set constraints for locating errors in CLP programs

WŁODZIMIERZ DRABENT∗, JAN MAŁUSZYŃSKI and PAWEŁ PIETRZAK

*Linköping University, Department of Computer and Information Science*
*S − 581 83 Linköping, Sweden*
*(e-mail: {*`wdr,jmz,pawpi`*}*`@ida.liu.se`*)*

## Abstract

This paper introduces a framework of parametric descriptive directional types for constraint logic programming (CLP). It proposes a method for locating type errors in CLP programs and presents a prototype debugging tool. The main technique used is checking correctness of programs w.r.t. type specifications. The approach is based on a generalization of known methods for proving correctness of logic programs to the case of parametric specifications. Set-constraint techniques are used for formulating and checking verification conditions for (parametric) polymorphic type specifications. The specifications are expressed in a parametric extension of the formalism of term grammars. The soundness of the method is proved and the prototype debugging tool supporting the proposed approach is illustrated on examples.

The paper is a substantial extension of the previous work by the same authors concerning monomorphic directional types.

## 1 Introduction

The objective of this work is to support development of CLP programs by a tool that checks correctness of a (partially developed) program wrt an approximate specification. Failures of such checks are used to locate fragments of the program which are potential program errors.

The specifications we work with extend the traditional concept of *directional type* for logic programs (see e.g. (Bronsart *et al.*, 1992)). Such a specification associates with every predicate a pair of sets that characterize, respectively, expected calls and successes of the predicate. Checking correctness of a logic program wrt directional types has been discussed by several authors (see e.g. (Aiken & Lakshman, 1994; Boye, 1996; Boye & Małuszyński, 1997; Charatonik & Podelski, 1998) and references therein). Their proposals can be seen as special cases of general verification methods of (Drabent & Małuszyński, 1988; Bossi & Cocco, 1989; Deransart, 1993). Technically, directional type checking consists in proving that the sets specified by given directional types of a program satisfy certain *verification conditions* constructed for this program. For directional types expressed as set constraints the verification conditions can also be expressed as set constraints and the check can be performed by set constraint techniques (see e.g. (Aiken & Lakshman, 1994)).

In this paper we propose an extension of directional types which addresses two issues:

- CLP programs operate on *constraint domains* while (pure) logic programs are restricted to one specific constraint domain which is the Herbrand universe. Directional types of a logic program characterize calls and successes of each predicate as sets of terms. This is not sufficient for CLP where manipulated data include *constraints* over non-Herbrand domains. To account for that we use a notion of *constrained term* where a constraint from a specific domain is attached to a non-ground term. We define the concept of directional type for CLP programs using sets of constrained terms.
- In logic programming, as well as in CLP, some procedures may be associated with families of directional types, rather than with single types. For example, typical list manipulation procedures may be used for lists with elements of any type and return lists with the elements of the same type. This is known as *parametric polymorphism* and can be described by a *parametric specification*, in our case by a *parametric directional type*. We extend the concept of partial correctness of CLP program to the case of parametric specifications and we give a sufficient condition for a program to be correct wrt a parametric specification. We apply this condition to correctness checking of CLP programs wrt parametric directional types, and for locating program errors. As shown by examples in Section 6, use of parametric specifications improves the possibility of locating errors.

The problem of checking of polymorphic directional types has been recently formulated in a framework of a formal calculus (Rychlikowski & Truderung, 2000; Rychlikowski & Truderung, 2001). As explained in Section 7.1 that approach is substantially different from ours.

A parametric specification can be seen as a family of (parameter-free) specifications. As mentioned above, our specifications refer to sets of constrained terms. The sufficient conditions for correctness can be formulated as set constraints, involving operations on the specified sets, such as projection, intersection and inclusion.

For constructing an automatic tool for checking correctness of specifications two questions have to be addressed:

- How to represent sets so that the necessary operations can be effectively performed,
- How to deal with parametric specifications.

The first problem was already discussed in (Drabent *et al.*, 2000b; Drabent *et al.*, 2000a), which extends our earlier work (Comini *et al.*, 1998; Comini *et al.*, 1999). We have chosen to represent sets of constrained terms by a simple extension of the formalism of discriminative term grammars, where sets of constrained terms are constructed from a finite collection of base sets. Term grammars (or equivalent formalisms) and set constraints have been used by many authors for specifying and inferring types for logic programs (see among others (Mishra, 1984; Frühwirth *et al.*, 1991; Dart & Zobel, 1992; Gallagher & de Waal, 1994; Aiken & Lakshman, 1994; Boye, 1996; Devienne *et al.*, 1997a; Charatonik & Podelski, 1998)). We show how the operations on discriminative term grammars can be extended to handle sets of constrained terms introduced by the extended discriminative term grammars.

A solution to the second problem is a main contribution of this paper. We derive it by showing how the approach of (Drabent *et al.*, 2000a) can be extended to the case of parametric specifications. (In our former work parametric grammars were used only in the user interface, to represent families of grammars.) First we have to give a new, more precise, presentation of that approach. We present a natural extension of the notion of partial correctness to the case of parametric specifications, so that the special case of parameterless specifications reduces to the notion used in our previous work. We introduce a concept of PED-grammar (parametric discriminative extended term grammar) as a formalism for specifying families of sets of constrained terms. We define operations on PED-grammars that make it possible to approximate results of the respective operations on members of the so defined families. We use them for checking correctness of programs wrt parametric directional types, and for locating potential errors.

If the verification conditions of a logic program are expressed as set constraints, it is possible to infer directional types that satisfy them. For example, the techniques of (Heintze & Jaffar, 1990a; Heintze & Jaffar, 1991) make it possible to construct a term grammar[1] describing the least model of the set constraints. The use of these techniques for program analysis in general was discussed in (Heintze, 1992).

On the other hand, it is possible to use abstract interpretation techniques to infer directional types of a program. Soundness of an abstract interpretation method can be justified by deriving it systematically from the verification conditions. An example of an abstract interpretation approach is (Janssens & Bruynooghe, 1992; Van Hentenryck *et al.*, 1995). A technique of (Gallagher & de Waal, 1994), similar to abstract interpretation, derives types in a form equivalent to discriminative term grammars. In (Drabent *et al.*, 2000a) we modified the latter technique to infer directional types for CLP programs. In this paper we present its further extension for inferring parametric directional types. We prove that this extension is sound in the sense that the program is correct wrt the inferred parametric types.

We use our technique of parametric type checking for locating errors in CLP programs. More precisely, we check correctness of a program wrt a parametric specification of directional types and we indicate fragments of clauses where the check of the verification conditions fails. However, CLP languages are often not typed so that programs do not include type specifications. Therefore our methodology does not require that the type specification is given a priori. The user decides a posteriori whether or not to type check a program, or its fragment.

The type specification is usually provided in a step-wise interactive way. At each stage of this process the program is checked against the fragment of the specification at hand. So incremental building of the specification is coupled together with locating errors. Even small fragments of the specification are often sufficient to locate (some) errors in the program. On the other hand, if no program errors have been located when the specification is completed then the program is correct (wrt the specification). Notice however that not every error message corresponds to the actual error in the program. That is why we call the error messages "warnings". This is due to using approximated specifications and to approximations made in the process of checking.

---

[1] In general this grammar is non-discriminative.

In the proposed methodology the process of type specification is preceded by static analysis which infers directional types of the program. The inferred types may provide indication that the program is erroneous. In this case the user may decide to start the process of specification and error location. The results of the type inference may facilitate it, as discussed below and in Section 6. Thus, in our methodology type inference plays only an auxiliary, though useful, role.

The methodology is supported by a prototype error locating tool. The present version of the tool works for a subset of the constraint programming language CHIP (Cosytec, 1998). However, it can be easily adapted for other CLP languages.

The structure of the tool is illustrated in Fig 1. The tool includes a type checker, a type inferencer and a specification editor. The tool has also a library of PED grammars. Among others, the library provides descriptions of often occurring types and specifications for built-in predicates. The specification of a program is introduced through the editor. It may refer to library grammars and/or to grammars provided by the user together with the checked program.

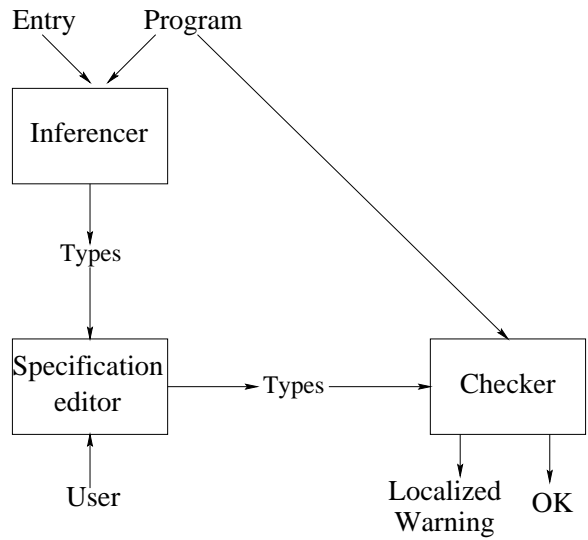

Fig. 1. The structure of the error locating tool

The input consists of a (possibly incomplete) CLP program and of an *entry declaration*. The latter is a parametric specification of intended (atomic) initial calls in terms of some PED grammar. In this way a family of sets is specified. Each member of the family is a different set of intended calls, corresponding to a different use of the program. The type inferencer constructs parametric directional types for all predicates of the program, thus providing a specification such that the input program is correct wrt to it. However, these types may not correspond to user intentions. This is due to program errors or to inaccuracy of type inference.

The intended types have to be provided by the user. They are introduced in a step-wise interactive manner. When providing the type of a predicate the user may first inspect the inferred type and accept it, or specify instead a different type. The tool monitors the process and immediately reports as an error any violation of the verification conditions for the so far introduced types.

While our approach makes it possible to locate some errors in CLP programs it should be clear that it is limited:

- It locates only type errors.
- Our types are based on discriminative regular grammars; the expressive power of this formalism is limited.
- To deal with constraints we extend this formalism from terms to constrained terms. However our treatment of constraints is rather crude. Roughly speaking, our formalism is able to define only a finite collection of sets of constraints (for any given variable). This limited approach lets us however find typical type bugs related to constraints. In our former work (Drabent & Pietrzak, 1998) we studied a more sophisticated (non parametric) type system for constrained terms. It seems however too complicated. Charatonik (1998) showed that a certain approach to approximating the semantics of CLP programs is bound to fail, as the resulting set constraints are undecidable.
- Correctness wrt parametric type specifications requires type correctness for all values of the type parameters. Thus only quite general sufficient conditions for correctness are possible. They however seem to work well on typical examples.

A usual question discussed in the literature is the theoretical worst case complexity of the proposed type checking and type inference algorithms. We show that our type checking algorithm for a clause is exponential wrt the number of variable repetitions. In our approach to locating errors type inference plays an auxiliary role and is implemented by an adaptation of the algorithm of (Gallagher & de Waal, 1994) with some ideas of (Mildner, 1999). While we prove soundness of this adaptation, we do not elaborate on the theoretical complexity issues, which by the way were not discussed by the authors of the algorithm. As concerns practical efficiency of our implementation, it turns out to be satisfactory on all examples we tried so far.

The main original contributions of the paper are:

- formulation of the concept of partial correctness of CLP programs wrt parametric specifications,
- a method for proving such correctness,
- a technique for checking of parametric directional types for CLP programs, based on this method,

- a prototype tool for locating program errors based on this technique.

The paper is organized as follows. Section 2 surveys some basic concepts on set constraints and constraint logic programs. Section 3 discusses the notion of correctness of a CLP program with respect to a specification, a sufficient condition for partial correctness and a technique for constructing approximations of program semantics. The main contributions of the paper are presented in the next sections. Section 4 introduces PED Grammars to be used as a parametric specification formalism for CLP programs. Section 5 introduces the notion of correctness wrt to a parametric specification and presents a method for proving such correctness. It shows how correctness can be effectively checked in case of parametric specifications provided as PED grammars. It also discusses how to construct a parametric specification of a given program. Finally it explains how program errors can be located by failures of the parametric correctness check. Section 6 discusses the prototype tool and illustrates its use on simple examples. Section 7 discusses relation to other work and presents conclusions.

This paper is an extended version of a less formal presentation of this work in (Drabent *et al.*, 2001).

## 2 Preliminaries

In this section we present some underlying concepts and techniques used in our approach. We introduce set constraints and term grammars. They are a tool to define sets of terms. Then we generalize them to define sets of constrained terms. The section is concluded with an overview of basic notions of constraint logic programming (CLP).

### 2.1 Set Constraints

This section surveys some basic notions and results on set constraints. We will extend them later to describe approximations of the semantics of CLP programs and to specify user expectations about behaviour of the developed programs.

We build set expressions from the alphabet consisting of: variables, function symbols (including constants), the intersection symbol $\cap$ and, for every variable $X$, the generalized projection symbol $^{-X}$.

A set expression is a variable, a constant, or it has a form $f(e_1, \ldots, e_n)$, $e_1 \cap e_2$, or $t^{-X}(e)$, where $f$ is an $n$-ary function symbol, $e, e_1, \ldots, e_n$ are set expressions, $t$ is a term and $X$ a variable. Set expressions built out of variables and function symbols (so including neither an intersection symbol nor a generalized projection symbol) are called *atomic*.

Set expressions are interpreted over the powerset of the Herbrand universe defined by a given alphabet. A valuation that associates sets of terms to variables extends to set expressions in a natural way: $\cap$ is interpreted as the intersection operation, each $n$-ary function symbol ($n \geq 0$) denotes the set construction operation

$$f(S_1, \ldots, S_n) = \{ f(t_1, \ldots, t_n) \mid t_i \in S_i, \ i = 1, \ldots, n \}$$

(for any sets $S_1, \ldots, S_n$ of ground terms) and symbol $t^{-X}$ denotes the generalized projection operation

$$t^{-X}(S) = \{\, X\theta \mid t\theta \in S, \ \theta \text{ is a substitution}, X\theta \text{ is ground} \,\}.$$

(for any term $t$, variable $X$ and set $S$ of ground terms)

Notice that we do not need special symbols for the projection operation and for the set of all terms. The latter is the value of $t^{-X}(S)$, where $X$ does not occur in $t$ and some instance of $t$ is in $S$. Projection, defined as $f_{(i)}^{-1}(S) = \{\, t_i \mid f(t_1, \ldots, t_n) \in S \,\}$, can be expressed as $f_{(i)}^{-1}(S) = f(X_1, \ldots, X_n)^{-X_i}(S)$.

Set expressions defined above are a proper subset of some classes of set expressions discussed in literature. In particular $t^{-X}(S)$ (where $X$ occurs in $t$) is a special case of the generalized membership expression of (Talbot *et al.*, 2000), in the notation of that paper it is $\{\, X \mid \exists_{-X} t \in S \,\}$. An (unnamed) operation more general than $t^{-X}$ has also been used in (Heintze & Jaffar, 1990b).

Our choice of the class of set expressions is guided by our application, which is parametric descriptive types for CLP programs. Later on we generalize set expressions to deal with sets of constrained terms (instead of terms) and to include parametric set expressions.

The set constraints we consider are of the form

$$Variable \ > \ Set \ expression$$

An interpretation of set constraints is defined by a valuation of variables as sets of ground terms. A model of a constraint is an interpretation that satisfies it when $>$ is interpreted as set inclusion $\supseteq$. Ordering on interpretations is defined by set inclusion: $I \leq I'$ iff $I(X) \subseteq I'(X)$ for every variable $X$. In such a case we will say that $I'$ approximates $I$. It can be proved (see for instance (Talbot *et al.*, 2000) and Proposition 2.9)  that a collection $G$ of such constraints is satisfiable and has the least model to be denoted $\mathcal{M}_G$. The value of a set expression $e$ in the least model of $G$ will be denoted by $[\![e]\!]_G$; the subscript may be omitted when it is clear from the context.

### *2.1.1 Term Grammars*

A finite set of constraints of the form

$$Variable \ > \ Atomic \ set \ expression$$

will be called *term grammar*. The least model of such a set of constraints can be obtained by assigning to each variable $X$ the set of all ground terms *derivable* from $X$ in this grammar. The derivability relation $\Rightarrow_G^*$ of a grammar $G$ is defined in a natural way: some occurrence of a variable $X$ in a given atomic set expression is replaced by a set expression $e$ such that $X > e$ is a constraint in $G$. Then $[\![X]\!]_G$ is the set of all ground terms derivable from $X$ in $G$.

A set $S$ is said to be defined by a grammar $G$ if there is a variable $X$ of $G$ such that $S = [\![X]\!]_G$. A grammar rule $X > t$ will be sometimes called a *rule for $X$*.

*Example 2.1*
For the following grammar the elements of $[\![List]\!]$ can be viewed as lists of bits.

$$List > nil \qquad\qquad B > 0$$
$$List > cons(B, List) \qquad\qquad B > 1$$

A pair $\langle X, G \rangle$ of a variable $X$ and a grammar $G$ uniquely determines the set $[\![X]\!]_G$ defined by the grammar; such a pair will be called a *set descriptor* (or a type descriptor). Sometimes we will say that $\langle X, G \rangle$ defines the set $[\![X]\!]_G$. By $\langle X \rangle_G$ we denote the collection of all rules of $G$ applicable in derivations starting from $X$.

We will mostly use a special kind of term grammars.

*Definition 2.2*
A term grammar is called *discriminative* iff

- each right hand side of a constraint is of the form $f(X_1, \ldots, X_n)$, where $X_1, \ldots, X_n$ are variables, and
- for a given variable $X$ and given $n$-ary function symbol $f$ there is at most one constraint of the form $X > f(\ldots)$

It should be mentioned that discriminative term grammars are just another view of deterministic top-down tree automata (Comon *et al.*, 1997). Variables of a grammar are states of an automaton, grammar derivations can be seen as computations of automata. Abandoning the second condition from Definition 2.2 leads to a strictly stronger formalism of non discriminative grammars equivalent to nondeterministic top-down tree automata.

We should explain our choice of the less powerful formalism of discriminative grammars. They seem to be sufficient to describe those sets which are usually considered to be types (Aiken & Lakshman, 1994) and also easier to understand for the user, which is important in our application. One of the goals of this work is enhancing term grammars with parameters. It seems reasonable to begin with a simpler formalism. We also want to find out to which extent a simpler formalism is sufficient in practice.

### 2.1.2  Operations on Term Grammars

The role of discriminative grammars is to define sets of terms. One needs to construct grammars describing the results of set operations on such sets. In this section we survey some operations on discriminative grammars, corresponding to set operations. A more formal presentation is given in Section 4 where we introduce a generalization of term grammars.

**Emptiness check.** A variable $X$ in a grammar $G$ will be called *nullable* if no ground term can be derived from $X$ in $G$. In other words, $[\![X]\!]_G = \emptyset$ iff $X$ is nullable in $G$. To check whether $[\![X]\!]_G = \emptyset$, one can apply algorithms for finding nullable symbols in context-free grammars. This can be done in linear time (Hopcroft *et al.*, 2001).

Let $G'$ be the grammar $G$ without the rules containing nullable symbols. Both grammars define the same sets, $[\![X]\!]_G = [\![X]\!]_{G'}$ for any variable $X$.

**Construction.** If $S_1, \ldots, S_n$ are defined by $\langle X_1, G_1 \rangle, \ldots, \langle X_n, G_n \rangle$, where $G_1, \ldots, G_n$ are discriminative grammars with disjoint sets of variables then the set $f(S_1, \ldots, S_n)$ is defined by $\langle X, G \rangle$ where $G$ is the discriminative grammar $\{X > f(X_1, \ldots, X_n)\} \cup G_1 \cup \ldots \cup G_n$ and $X$ is a new variable, not occurring in $G_1, \ldots, G_n$.

**Intersection.** Given sets $S$ and $T$ defined by discriminative grammars $G_1$ and $G_2$ we construct a discriminative grammar $G$ such that $S \cap T$ is defined by $G$. Without loss of generality we assume that $G_1$ and $G_2$ have no common variables. The variables of $G$ correspond to pairs $(X, Y)$ where $X$ is a variable of $G_1$ and $Y$ is a variable of $G_2$. They will be denoted $X \dot{\cap} Y$. The notation reflects the intention that $[\![(X, Y)]\!]_G = [\![X]\!]_{G_1} \cap [\![Y]\!]_{G_2}$.

Now $G$ is defined as the set of all rules

$$X \dot{\cap} Y > f(X_1 \dot{\cap} Y_1, \ldots, X_n \dot{\cap} Y_n)$$

such that there exist a rule $X > f(X_1, \ldots, X_n)$ in $G_1$ and a rule $Y > f(Y_1, \ldots, Y_n)$ in $G_2$. Notice that for given $f$ at most one rule of this form may exist in each of the grammars. Thus $G$ is discriminative. It is not difficult to prove that $[\![(X, Y)]\!]_G$ is indeed the intersection of $[\![X]\!]_{G_1}$ and $[\![Y]\!]_{G_2}$.

We have $S = [\![X]\!]_{G_1}$ for some $X$ of $G_1$ and $T = [\![Y]\!]_{G_2}$ for some $Y$ of $G_2$, hence $S \cap T$ is defined by $G$. Notice that $G$ may contain nullable symbols even if $G_1, G_2$ do not.

*Example 2.3*

Consider two grammars

$$
\begin{array}{ll}
G_1: \quad X > a & \qquad G_2: \quad Y > a \\
\qquad X > f(Z, Z) & \qquad \qquad Y > f(E, Y) \\
\qquad Z > f(X, X) & \qquad \qquad E > a \\
\qquad Z > b & \qquad \qquad E > b \\
\qquad Z > g(Z) & \qquad \qquad E > h(E)
\end{array}
$$

The grammar defining the intersections of the sets defined by $G_1, G_2$ is

$$
\begin{array}{l}
G: \quad X \dot{\cap} Y > a \\
\qquad X \dot{\cap} Y > f(Z \dot{\cap} E, Z \dot{\cap} Y) \\
\qquad Z \dot{\cap} Y > f(X \dot{\cap} E, X \dot{\cap} Y) \\
\qquad X \dot{\cap} E > a \\
\qquad Z \dot{\cap} E > b
\end{array}
$$

**Union.** It is well known that the union of sets defined by discriminative grammars may not be definable by a discriminative grammar; take for example the sets $\{f(a, b)\}$ and $\{f(c, d)\}$. Given sets $S$ and $T$ defined by discriminative grammars $G_1$ and $G_2$ we construct now a discriminative grammar $G$ defining a superset of $S \cup T$.

Without loss of generality we assume that $G_1$ and $G_2$ have no common variables. The variables of $G$ correspond to pairs $(X, Y)$ where $X$ is a variable of $G_1$ and $Y$ is a variable of $G_2$. They will be denoted $X \dot{\cup} Y$. The notation reflects the intention that $[\![X]\!]_{G_1} \cup [\![Y]\!]_{G_2} \subseteq [\![(X, Y)]\!]_G$.

Now $G$ consists of the rules of $G_1$, the rules of $G_2$ and of the least set of rules which can be constructed as follows:

- If $X > f(X_1, \ldots, X_n)$ is in $G_1$ and $Y > f(Y_1, \ldots, Y_n)$ is in $G_2$ then $X \dot\cup Y > f(X_1 \dot\cup Y_1, \ldots, X_n \dot\cup Y_n)$ is in $G$,
- If $X > f(X_1, \ldots, X_n)$ is in $G_1$ and no rule $Y > f(Y_1, \ldots, Y_n)$ is in $G_2$ then $X \dot\cup Y > f(X_1, \ldots, X_n)$ is in $G$,
- If no rule $X > f(X_1, \ldots, X_n)$ is in $G_1$ and $Y > f(Y_1, \ldots, Y_n)$ is in $G_2$ then $X \dot\cup Y > f(Y_1, \ldots, Y_n)$ is in $G$

It is not difficult to see that the obtained grammar $G$ is discriminative, and that $[\![X \dot\cup Y]\!]_G$ is indeed a superset of the union of $[\![X]\!]_{G_1}$ and $[\![Y]\!]_{G_2}$. If the first case is not involved in the construction the result is the union of these sets. If $G_1, G_2$ do not contain nullable symbols then $[\![X \dot\cup Y]\!]_G$ is the tuple-distributive closure of $[\![X]\!]_{G_1} \cup [\![Y]\!]_{G_2}$, i.e. the least set definable by a discriminative grammar and including $[\![X]\!]_{G_1} \cup [\![Y]\!]_{G_2}$. (We skip a proof of this fact, we do not use it later). So we are able to obtain the best possible approximation of the union by a discriminative grammar.

*Example 2.4*
The singleton sets $\{f(a, b)\}$ and $\{f(c, d)\}$ can be defined by the grammars:

$$G_1 \colon X > f(A, B),\ A > a,\ B > b \qquad G_2 \colon Y > f(C, D),\ C > c,\ D > d.$$

Applying the construction we obtain additional rules:

$$X \dot\cup Y > f(A \dot\cup C, B \dot\cup D) \qquad A \dot\cup C > a \qquad B \dot\cup D > b$$
$$\qquad\qquad\qquad\qquad A \dot\cup C > c \qquad B \dot\cup D > d$$

**Set inclusion** Given sets $S$ and $T$ defined by discriminative grammars it is possible to check $S \subseteq T$ by examination of the defining grammars.

By the assumption $S = [\![X]\!]_{G_1}, T = [\![Y]\!]_{G_2}$ for some discriminative grammars $G_1, G_2$ and some variables $X, Y$. We assume without loss of generality that $G_1, G_2$ do not contain nullable symbols. (Otherwise the nullable symbols may be removed as justified previously).

It follows from the definition of the set defined by term grammar that $[\![X]\!]_{G_1} \subseteq [\![Y]\!]_{G_2}$ iff for every rule of the form $X > f(X_1, \ldots, X_n)$ in $G_1$ there exists a rule $Y > f(Y_1, \ldots, Y_n)$ in $G_2$ and $[\![X_i]\!]_{G_1} \subseteq [\![Y_i]\!]_{G_2}$ for $i = 1, \ldots, n$. This corresponds to a recursive procedure where a check for $X, Y$ corresponds to comparison of function symbols in the defining rules for $X$ and $Y$, which may cause a failure, and a recursive call of a finite number of such checks. The check performed once for a given pair of variables need not be repeated. As the grammar is finite there is a finite number of pairs of variables so that the check will terminate.

For a formal description of the algorithm and a correctness proof see Section 4.4.5 where a more general inclusion check algorithm is presented.

*Example 2.5*
The following example illustrates inclusion checking. It shows that the set of non-empty bit lists with even length is a subset of the set of unrestricted lists which allow a more general kind of elements. Both sets are described by discriminative grammars.

$$
\begin{array}{ll}
S > cons(B, Odd) & List > nil \\
Odd > cons(B, Even) & List > cons(E, List) \\
Even > nil & E > 0 \\
Even > cons(B, Odd) & E > 1 \\
B > 0 & E > s(E) \\
B > 1 &
\end{array}
$$

We check inclusion $[\![S]\!] \subseteq [\![List]\!]$. We show steps of this process. Each step will be characterized by three items: the checked pair of variables, the function symbols in their defining rules, the set of pairs to be checked after this step.

$$
\begin{array}{lll}
(S, List) & (\{cons\}, \{nil, cons\}) & \{\,(B, E), (Odd, List)\,\} \\
(B, E) & (\{0, 1\}, \{0, 1, s\}) & \{\,(Odd, List)\,\} \\
(Odd, List) & (\{cons\}, \{nil, cons\}) & \{\,(Even, List)\,\} \\
(Even, List) & (\{nil, cons\}, \{nil, cons\}) & \emptyset
\end{array}
$$

**Generalized projection.** Assume that $S = [\![Y]\!]_G$ is defined by a discriminative grammar $G$. We show that $t^{-X}(S)$ is defined by a discriminative grammar.

Consider a term $t$ and a mapping $\xi(t, G, Y)$ assigning a variable $V_u$ of $G$ to each subterm occurrence $u$ of $t$, such that $V_t$ is $Y$ and if $u = f(u_1, \ldots, u_n)$ $(n \geq 0)$ then there exists a rule $V_u > f(V_{u_1}, \ldots, V_{u_n})$ in $G$. So for instance in Example 2.5, taking $t = cons(s(X), Z)$ and $Y = List$ results in $V_t = List$, $V_{s(X)} = E$, $V_Z = List$, $V_X = E$. If such a mapping exists then it is unique, as the grammar contains at most one rule $V > f(\ldots)$ for given $V, f$.

The mapping can be found by an obvious algorithm. It traverses $t$ top-down and for each occurrence $u$ of a non-variable subterm it finds the unique rule $V_u > f(V_{u_1}, \ldots, V_{u_n})$. The rule determines the variables $V_{u_1}, \ldots, V_{u_n}$ corresponding to the greatest proper subterms of $u$. If such a rule does not exist, mapping $\xi(t, G, Y)$ does not exist. The starting point is $u = t$ and $V_u = Y$.

Notice that if $t\theta \in S$ then $\xi(t, G, Y)$ exists and $u\theta \in [\![V_u]\!]_G$ for each subterm occurrence $u$ in $t$. Hence $X\theta \in [\![V_{X^i}]\!]_G$ for each occurrence $X^i$ of $X$ in $t$. Thus $t^{-X}(S) \subseteq \bigcap_i [\![V_{X^i}]\!]_G$. (If $X$ does not occur in $t$ then $\bigcap_i [\![V_{X^i}]\!]_G$ denotes the Herbrand universe.) On the other hand, assume that $\xi(t, G, Y)$ exists and for each variable $Z$ of $t$ there exists a term $u_Z$ such that $u_Z \in [\![V_{Z^i}]\!]_G$ for each occurrence $Z^i$ of $Z$ in $t$. Then $t\theta \in S$, where $\theta = \{\, Z/u_Z \mid Z \text{ occurs in } t \,\}$. Thus if $\xi(t, G, Y)$ exists and $\bigcap_i [\![V_{Z^i}]\!]_G$ is nonempty for each $Z$ then

$$
t^{-X}(S) = \bigcap_i [\![V_{X^i}]\!]_G \, .
$$

Otherwise $t^{-X}(S) = \emptyset$.

Applying algorithms described previously, we can construct for each $Z$ a distributive grammar $G_Z$ defining $[\![Z']\!]_{G_Z} = \bigcap_i [\![V_{Z^i}]\!]_G$ and check this set for emptiness. This provides an algorithm which, given $G, Y, t$, produces for each $X$ occurring in $t$ a discriminative grammar $G_X$ and a variable $X'$ such that $t^{-X}(S) = [\![X']\!]_{G_X}$.

An algorithm similar to the presented above is used in the implementation of (Gallagher & de Waal, 1994), it is however only superficially described in that paper.

## 2.2 Specifying sets of constrained terms

Set constraints and term grammars are formalisms for defining subsets of the Herbrand universe. This is not sufficient for the purposes of CLP. We use a CLP semantics based on the notion of a constrained expression. The goal of this section is generalizing discriminative term grammars to a mechanism of defining sets of constrained terms.

### 2.2.1 Constrained expressions

CLP programs operate on constraint domains. A *constraint domain* is defined by providing a finite signature (of predicate and function symbols) and a structure $\mathcal{D}$ over this signature.[2] Predicate symbols of the signature are divided into *constraint predicates* and *non-constraint predicates*. The former have a fixed interpretation in $\mathcal{D}$, the interpretation of the latter is defined by programs. All the function symbols have a fixed interpretation, they are interpreted as constructors. So the elements of $\mathcal{D}$ can be seen as (finite) terms built from some elementary values and the constant symbols by means of constructors. That is why we will often call them $\mathcal{D}$-terms. In CLP some function symbols have also other meaning (like + denoting addition in CLP over integers). This meaning is employed only in the semantics of constraint predicates.

We treat function symbols as constructors, because this happens in the semantics of most CLP languages, like CHIP or SICStus Prolog (Cosytec, 1998; SICS, 1998). They use syntactic unification. For instance, in CLP over integers, terms like $1 + 3$, $2 + 2$, $1 * 4$, $4$ are (pairwise) not unifiable. Only the constraint predicates recognize their numerical values. So $2 + 2 \ \#= \ 1 * 4$ succeeds and $2 + 2 \ \#> \ 3 * 4$ fails (where $\#=$, $\#>$ are constraint predicates of, respectively, arithmetical equality and comparison).

By a *constraint* we mean an atomic formula with a constraint predicate, $c_1 \wedge c_2$, $c_1 \vee c_2$, or $\exists X c_1$, where $c_1$ and $c_2$ are constraints and $X$ is a variable. We will often write $c_1, c_2$ for $c_1 \wedge c_2$. The fact that a constraint $c$ is true for every variable valuation will be denoted by $\mathcal{D} \models c$.

The Herbrand domain of logic programming is generalized to the constraint domain $\mathcal{D}$ of CLP. Analogical generalization of non ground atoms and terms are constrained expressions.

*Definition 2.6*
A *constrained expression* (atom, term) is a pair $c \, [\!] \, E$ of a constraint $c$ and an expression $E$ such that each free variable of $c$ occurs (freely) in $E$.

---

[2]  Sometimes we slightly abuse the notation and use $\mathcal{D}$ to denote the carrier of $\mathcal{D}$.

A $c \parallel E$ with some free variable of $c$ not occurring in $E$ will be treated as an abbreviation for $(\exists \ldots c) \parallel E$, where all variables of $c$ not occurring in $E$ are existentially quantified,

*Definition 2.7*
A constrained expression $c' \parallel E'$ is an *instance* of a constrained expression $c \parallel E$ if $c'$ is satisfiable in $\mathcal{D}$ and there exists a substitution $\theta$ such that $E' = E\theta$ and $\mathcal{D} \models c' \rightarrow c\theta$ ($c\theta$ means here applying $\theta$ to the free variables of $c$, with a standard renaming of the non-free variables of $c$ if a conflict arises).

If $c \parallel E$ is an instance of $c' \parallel E'$ and vice versa then $c \parallel E$ is a *variant* of $c' \parallel E'$.

By the *instance-closure* $cl(E)$ of a constrained expression $E$ we mean the set of all instances of $E$. For a set $S$ of constrained expressions, its instance-closure $cl(S)$ is defined as $\bigcup_{E \in S} cl(E)$.

Note that, in particular, $c\theta \parallel E\theta$ is an instance of $c \parallel E$ and that $c' \parallel E$ is an instance of $c \parallel E$ whenever $\mathcal{D} \models c' \rightarrow c$, provided that $c\theta$ and, respectively, $c'$ are satisfiable. The relation of being an instance is transitive. (Take an instance $c' \parallel E\theta$ of $c \parallel E$ and an instance $c'' \parallel E\theta\sigma$ of $c' \parallel E\theta$. As $\mathcal{D} \models c'' \rightarrow c'\sigma$ and $\mathcal{D} \models c' \rightarrow c\theta$, we have $\mathcal{D} \models c'' \rightarrow c\theta\sigma$). Notice also that if $c$ is not satisfiable then $c \parallel E$ does not have any instance (it is not an instance of itself).

We will often not distinguish $E$ from $true \parallel E$ and from $c \parallel E$ where $\mathcal{D} \models \forall c$. Similarly, we will also not distinguish $c \parallel E$ from $c' \parallel E$ when $c$ and $c'$ are equivalent constraints ($\mathcal{D} \models c \leftrightarrow c'$).

*Example 2.8*
$a + 7$, $Z + 7$, $1 + 7$ are instances of $X + Y$, but 8 is not.

$f(X){>}3 \parallel f(X){+}7$ is an instance of $Z{>}3 \parallel Z{+}7$, which is an instance of $Z + 7$, provided that constraints $f(X){>}3$ and $Z{>}3$, respectively, are satisfiable.

Assume a numerical domain with the standard interpretation of symbols. Then $4 + 7$ is an instance of $X{=}2{+}2 \parallel X{+}7$ (but not vice versa), the latter is an instance of $Z{>}3 \parallel Z{+}7$.

Consider CLP(FD) (CLP over finite domains, (Van Hentenryck, 1989)). A domain variable with the domain $S$, where $S$ is a finite set of natural numbers, can be represented by a constrained variable $X{\in}S \parallel X$ (with the expected meaning of the constraint $X{\in}S$).

### 2.2.2 Extended Set Constraints

We use a semantics for CLP which is based on constrained atoms/terms. To approximate such semantics we generalize term grammars to describe instance-closed sets of constrained terms. In discussing grammars and the generated sets, we will not distinguish between predicate and function symbols, and between atoms and terms.

For a given constraint domain $\mathcal{D}$, we introduce some *base sets* of constrained terms. We require that base sets are instance-closed. Following (Dart & Zobel, 1992) we extend the alphabet of set constraints by *base symbols* interpreted as base sets. Each base symbol $b$ has a fixed corresponding set $\llbracket b \rrbracket$ of constrained terms, $\llbracket b \rrbracket \neq \emptyset$. We require that the alphabet of base symbols is finite. We assume that there is a base symbol $\top$ for which $\llbracket \top \rrbracket$ is the set of all constrained terms over given $\mathcal{D}$. Usually no other base sets contain (constrained) terms with (non constant) function symbols.

For instance in CLP over finite domains (Van Hentenryck, 1989), $\mathcal{D}$ contains terms built of symbols and integer numbers. The base sets we use for this domain are, apart from $\llbracket \top \rrbracket$, denoted by base symbols $nat$, $neg$, $anyfd$. They correspond to, respectively, the natural numbers, the negative integers and finite domain variables. The latter are represented as constrained variables of the form $X \in S \, \rrbracket \, X$, where $S$ is a finite set of natural numbers. Due to the closedness requirement, $\llbracket anyfd \rrbracket$ contains also the natural numbers.

An *extended set expression* is an expression built out of variables, base symbols, function symbols (including constants), $\cap$ and the generalized projection symbols. Extended set expressions are interpreted as instance-closed sets of constrained terms. In the context of extended set expressions, a *valuation* is a mapping assigning instance-closed sets of constrained terms to variables.[3]

The construction and generalized projection operation for (instance closed) sets of constrained terms are defined as

$$f(S_1, \ldots, S_n) = cl(\{\, c_1, \ldots, c_n \, \rrbracket \, f(t_1, \ldots, t_n) \mid c_i \, \rrbracket \, t_i \in S_i, \ i = 1, \ldots, n \,\}),$$

$$t^{-X}(S) = \{\, c \, \rrbracket \, X\theta \mid c \, \rrbracket \, t\theta \in S, \ \text{for some substitution } \theta \,\},$$

for instance-closed sets $S, S_1, \ldots, S_n$, a function (or predicate) symbol $f$, a term (or an atom) $t$ and a variable $X$. Notice that $f(S_1, \ldots, S_n), t^{-X}(S)$ are instance-closed. A valuation, together with a fixed valuation of base symbols, extends in a natural way to extended set expressions. So if sets $S_1, \ldots, S_n$ are values of expressions $e_1, \ldots, e_n$ then the value of $f(e_1, \ldots, e_n)$ is $f(S_1, \ldots, S_n)$. For a ground extended set expression $t$ its value will be denoted by $\llbracket t \rrbracket$.

Extended set expressions can be used to construct set constraints and grammars. We consider *extended set constraints* of the form $X > t$, where $X$ is a variable and $t$ an extended set expression. An *extended term grammar* is a set of constraints (often called rules) of the form $X > t$, where $t$ is an atomic set expression (i.e. one built out of variables, the base symbols and the function symbols, including constants).

A *model* of a set $\mathcal{C}$ of extended set constraints is a valuation $I$, under which $I(X) \supseteq I(t)$ for each constraint $X > t$ of $\mathcal{C}$.

*Proposition 2.9*
Any set $\mathcal{C}$ of extended set constraints has the least model.

---

[3]  Notice that we have two different languages using variables: the language of set expressions (and of set constraints and grammars), with variables ranging over sets of constrained terms, and the language of constrained terms with variables ranging over a specific constraint domain. In this paper we use the same notation for both kinds of variables. This should cause no confusion, the kind of a variable is determined by the context.

*Proof*

We show that the set of models of $\mathcal{C}$ is nonempty and that their greatest lower bound is a model of $\mathcal{C}$.

$I$ assigning to each variable the set $\llbracket \top \rrbracket$ of all constrained terms is a model of any extended set constraint.

The greatest lower bound of a set $\mathcal{I}$ of valuations is a valuation $\bigcap \mathcal{I}$ such that $(\bigcap \mathcal{I})(X) = \bigcap \{ I(X) \mid I \in \mathcal{I} \}$, for any variable $X$.

Let $\circ$ be a construction operation, a generalized projection operation or $\cap$. Let $k$ be its arity. For $i = 1, \ldots, k$, let $\mathcal{S}_i$ be a set of instance closed sets of constrained terms. We have

$$\circ (\bigcap \mathcal{S}_1, \ldots, \bigcap \mathcal{S}_k) \ \subseteq \ \bigcap \{ \circ (S_1, \ldots, S_k) \mid S_1 \in \mathcal{S}_1, \ldots, S_k \in \mathcal{S}_k \}.$$

(We do not need here to show equality). Hence for any extended set expression $t$ and any set $\mathcal{I}$ of valuations

$$(\bigcap \mathcal{I})(t) \ \subseteq \ \bigcap \{ I(t) \mid I \in \mathcal{I} \},$$

by induction on the structure of $t$. Hence if each element of $\mathcal{I}$ is a model of an extended set constraint $X > t$ then $\bigcap \mathcal{I}$ is a model of $X > t$, as $(\bigcap \mathcal{I})(X) = \bigcap \{ I(X) \mid I \in \mathcal{I} \} \supseteq \bigcap \{ I(t) \mid I \in \mathcal{I} \} \supseteq (\bigcap \mathcal{I})(t)$. Thus if $\mathcal{I}$ is the set of models of $\mathcal{C}$ then $\bigcap \mathcal{I}$ is a model of $\mathcal{C}$, hence the least model. $\qquad \square$

*Definition 2.10*

The set defined by a variable $X$ in an extended term grammar $G$ is

$$\llbracket X \rrbracket_G = \{ c \,\llbracket\, u \mid c \,\llbracket\, u \in \llbracket t \rrbracket, \ X \Rightarrow_G^* t \text{ and no variable occurs in } t \}$$

where the derivability relation $\Rightarrow_G^*$ is defined as for term grammars.

Notice that we avoid confusion between the variables of grammars and the variables of constrained terms. The former occur in derivations, which end with ground terms built of function symbols (including constants) and of base symbols. The latter appear later on as a result of evaluation of base symbols in these ground terms.

The notation $\llbracket X \rrbracket_G$ is justified here by the following property.

*Proposition 2.11*

Let $G$ be an extended term grammar and $I$ the interpretation such that $I(X) = \llbracket X \rrbracket_G$ for each variable $X$. Then $I$ is the least model of $G$.

*Proof*

Consider a variable $X$ and a constrained term $c \,\llbracket\, s \in \llbracket X \rrbracket_G$. So there exists a derivation $X \Rightarrow_G^* t$ such that $c \,\llbracket\, s \in \llbracket t \rrbracket$. By induction on the length of the derivation, for any model $J$ of $G$, $\llbracket t \rrbracket \subseteq J(X)$. Thus $I(X) \subseteq J(X)$. Hence $I \leq J$. $\qquad \square$

*Definition 2.12*
An *extended discriminative term grammar* $G$ is a finite set of rules of the form

$$X > f(X_1, \ldots, X_n) \qquad \text{or} \qquad X > b$$

where $f$ is an $n$-ary function symbol ($n \geq 0$), $X, X_1, \ldots, X_n$ are variables and $b$ is a base symbol. Additionally, for each pair of rules $X > t_1$ and $X > t_2$ in $G$ the sets $[\![t_1^\top]\!]$ and $[\![t_2^\top]\!]$ are disjoint (where $u^\top$ stands for $u$ with each occurrence of a variable replaced by $\top$).

So no two rules $X > f(\vec{X})$, $X > f(\vec{Y})$ may occur in such a grammar. The same for $X > b$, $X > b'$ where $b, b'$ are base symbols and $[\![b]\!] \cap [\![b']\!] \neq \emptyset$. If a discriminative grammar contains $X > f(\vec{X})$ and $X > b$ then no (constrained term) with the main symbol $f$ occurs in $[\![b]\!]$. If the grammar contains $X > \top$ then it is the only rule for $X$.

The question is how to represent/approximate by such grammars the results of set operations for sets represented by such grammars, and how to check inclusion for such sets. We address these questions under some additional restrictions on base sets, which seem to be observed in base domains of CLP languages. We require that:

*Requirement 2.13*
- For any base symbol $b$ different from $\top$, $f_{(i)}^{-1}([\![b]\!]) = \emptyset$ for every $f, i$. (So $[\![b]\!]$ does not contain elements of the form $c \, [\!] \, f(\vec{t})$, for any non constant $f$.)
- For each pair $b_1, b_2$ of distinct base symbols the base sets $[\![b_1]\!]$, $[\![b_2]\!]$ are either disjoint or one is a subset of the other. Moreover $[\![b_1]\!] \neq [\![b_2]\!]$.

The number of base symbols is finite. Their interpretation is fixed. We can construct a table showing, for each pair $b_1, b_2$ of base symbols, whether $[\![b_1]\!] \cap [\![b_2]\!] = \emptyset$, $[\![b_1]\!] \subseteq [\![b_2]\!]$ or $[\![b_2]\!] \subseteq [\![b_1]\!]$.

Now, the operations on grammars of Section 2.1.1 can be easily extended. Each of them traverses the rules in the argument grammars. Eventually we may reach a point when a base symbol is encountered instead of a constant. These cases are handled in a rather obvious way, using the table described above. Similarly as for discriminative term grammars, one obtains approximation of the union and exact intersection, generalized projection and construction.

We postpone a formal presentation to Section 4.4, where we deal with a generalization of grammars discussed here.

*Example 2.14*
Consider CLP(FD) (Van Hentenryck, 1989). The following discriminative extended grammars describe, respectively, integer lists and lists of finite domain variables (possibly instantiated to natural numbers):

$$
\begin{array}{ll}
Li > nil & Lfd > nil \\
Li > cons(Int, Li) & Lfd > cons(A, Lfd) \\
Int > nat & A > anyfd \\
Int > neg &
\end{array}
$$

Knowing that $[\![nat]\!] \subseteq [\![anyfd]\!]$ we can apply the intersection operation to obtain a grammar defining $[\![Li\,]\!] \cap [\![Lfd]\!]$:

$$Li \mathbin{\dot\cap} Lfd > nil$$
$$Li \mathbin{\dot\cap} Lfd > cons(Int \mathbin{\dot\cap} A, Li \mathbin{\dot\cap} Lfd)$$
$$Int \mathbin{\dot\cap} A > nat$$

The treatment of constraints by the formalism of extended term grammars is rather rough. It stems from a small number of fixed base sets of constrained terms. They are subject to a rather restrictive Requirement 2.13, which is necessary to simplify operations on grammars. In our former work (Drabent & Pietrzak, 1998) we discussed a richer system of regular sets of constrained terms. It can be seen as also allowing base sets of the form $cl(\{c \mathbin{[\![} x\})$, where the set of ground terms satisfying constraint $c$ is regular. This results in substantially more complicated algorithms for grammar operations. According to our experience the simple type system presented in this paper seems sufficient.

### 2.3 Constraint Logic Programming

We consider CLP programs executed with the Prolog selection rule (LD-resolution) and using syntactic unification in the resolution steps. In CLP with syntactic unification, function symbols occurring outside of constraints are treated as constructors. So, for instance in CLP over integers, the goal $p(4)$ fails with the program $\{p(2{+}2)\leftarrow\}$ (but the goal $p(X{+}Y)$ succeeds). Terms 4 and $2{+}2$ are treated as not unifiable despite having the same numerical value. Also, a constraint may distinguish such terms. For example in many constraints of CHIP, an argument may be a natural number (or a "domain variable") but not an arithmetical expression. Resolution based on syntactic unification is used in many CLP implementations, for instance in CHIP and in SICStus (SICS, 1998).

We are interested in *calls* and *successes* of program predicates in computations of the program. Both calls and successes are constrained atoms. A precise definition is given below taking a natural generalization of LD-derivation as a model of computation.

An *LD-derivation* is a sequence $G_0, C_1, \theta_1, G_1, \ldots$ of goals, input clauses and mgu's (similarly to (Lloyd, 1987)). A goal is of the form $c \mathbin{[\![} A_1, \ldots, A_n$, where $c$ is a constraint and $A_1, \ldots, A_n$ are atomic formulae (including atomic constraints). For a goal $G_{i-1} = c \mathbin{[\![} A_1, \ldots, A_n$, where $A_1$ is not a constraint, and a clause $C_i = H \leftarrow B_1, \ldots, B_m$, the next goal in the derivation is $G_i = (c \mathbin{[\![} B_1, \ldots, B_m, A_2, \ldots, A_n)\theta_i$ provided that $\theta_i$ is an mgu of $A_1$ and $H$, $c\theta_i$ is satisfiable and $G_{i-1}$ and $C_i$ do not have common variables. If $A_1$ is a constraint then $G_i = c, A_1 \mathbin{[\![} A_2, \ldots, A_n$ ($\theta_i = \epsilon$ and $C_i$ is empty) provided that $c, A_1$ is satisfiable.

For a goal $G_{i-1}$ as above we say that $c \mathbin{[\![} A_1$ is a *call* (of the derivation). The call succeeds in the first goal of the form $G_k = c' \mathbin{[\![}(A_2, \ldots, A_n)\rho$ (where $k \geq i$, $\rho = \theta_i \cdots \theta_k$) of the derivation. The *success* corresponding (in the derivation) to the call above is $c' \mathbin{[\![} A_1\rho$. For example, $X{\in}\{1, 2, 3, 4\} \mathbin{[\![} p(X, Y)$ and $X{\in}\{1, 2, 4\} \mathbin{[\![} p(X, 7)$ is a possible pair of a call and a success for $p$ defined by $p(X, 7) \leftarrow X \neq 3$.

Notice that in this terminology constraints succeed immediately. If $A$ is a constraint then the success of call $c \,[\!]\, A$ is $c, A \,[\!]\, A$, provided $c, A$ is satisfiable. So we do not treat constraints as delayed; we abstract from internal actions of the constraint solver.

The *call-success semantics* of a program $P$, for a set of initial goals $\mathcal{G}$, is a pair $CS(P, \mathcal{G}) = (C, S)$ of sets of constrained atoms: the set of calls and the set of successes that occur in the LD-derivations starting from goals in $\mathcal{G}$. We assume without loss of generality that the initial goals are atomic.

So the call-success semantics describes precisely the calls and the successes in the considered class of computations of a given program. The question is whether this set includes "wrong" elements, unexpected by the user. To require a precise description of user expectations is usually not realistic. On the other hand, it may not be difficult to provide an approximate description $Spec = (C', S')$ where $C'$ and $S'$ are sets of constrained atoms such that every expected call is in $C'$ and every expected success is in $S'$.

*Definition 2.15*

A program $P$ with the set of initial goals $\mathcal{G}$ is *partially correct* w.r.t. $Spec = (C', S')$ iff $C \subseteq C'$ and $S \subseteq S'$, where $(C, S) = CS(P, \mathcal{G})$ is the call-success semantics of $P$ and $\mathcal{G}$.

$P$ is partially correct w.r.t. $Spec = (C', S')$ iff $P$ with $C'$ as the set of initial goals is partially correct w.r.t. $Spec$.

We will usually omit the word "partially".

To avoid substantial technical difficulties, we will consider only specifications that are closed under instantiation. This means that whenever set $C'$ (or $S'$) contains a constrained atom $c \,[\!]\, A$ then it contains all its instances.

In Section 5 we introduce parametric specifications, discuss a more precise semantics and generalize accordingly the notion of program correctness.

Our discussion of CLP semantics has been carried on under an assumption that the constraint solver is complete. Thus it is able to recognize all unsatisfiable constraints. However actual solvers are usually incomplete. As a result, goals with unsatisfiable constraints may appear in derivations. But the set of solutions represented by all answers of an incomplete solver is the same as the set of solutions represented by all answers of a complete solver. Thus, if our type checking technique indicates (possibility of) the existence of a wrong answer, beyond those characterized by a specification, then this answer will also be obtained with an incomplete solver. Thus the assumption on completeness of the solver is only a technicality needed for formal development of the method, which is also applicable in the case of incomplete solvers.

A specification describes calls and successes of all the predicates of a program, including the constraint predicates. As the semantics of constraints is fixed for a given programming language, their specification is fixed too. In our system it is kept in a system library and is not intended to be modified by the user. (The same happens for other built-in predicates of the language.) This fixed part of the specification may not permit some constrained atoms as procedure calls; such calls are not allowed in the language and result in run-time errors.[4]

*Example 2.16*
To illustrate the treatment of constraint predicates by specifications, assume that a CLP(FD) language has a constraint $\in$, which describes membership in a finite domain. Assume that invoking $\in(X, S)$ with $S$ not being a list of natural numbers is an error. This should be reflected by the specifications of all programs using $\in$. In any such specification $Spec = (Pre, Post)$, a call of the form $c \,\square\, \in(X, S)$ is in $Pre$ iff $S$ is such a list. If such a call succeeds, $X$ must be a finite domain variable or a natural number. We may thus require that $c \,\square\, \in(X, S)$ is in $Post$ iff $S$ is a list of natural numbers and $c \,\square\, X$ is in $[\![\,anyfd\,]\!]$.

The following definition provides a condition assuring that a specification correctly approximates successes of constraint predicates.

*Definition 2.17*
We say that a specification $(Pre, Post)$ *respects constraints* if $c, A \,\square\, A \in Post$ whenever $c \,\square\, A \in Pre$ and $c, A$ is satisfiable (for any constraint $c$ and atomic constraint $A$). This is equivalent to

$$\{\, c, A \,\square\, A \mid c, A \text{ is satisfiable} \,\} \cap Pre \subseteq Post$$

as $Pre$ is closed under instantiation.

## 3 Partial correctness of programs

In this section we present a verification condition for partial correctness of CLP programs. Then we express it by means of set constraints and show how to perform correctness checking and how to compute a specification approximating the call-success semantics of a program.

### *3.1 Verification condition*

A sufficient condition for such correctness of logic programs was given in (Drabent & Małuszyński, 1988). For specifications which are closed under substitution the condition is simpler (Bossi & Cocco, 1989), (Apt, 1997). Generalizing the latter for constraint logic programs we obtain:

*Proposition 3.1*
Let $P$ be a CLP program, $\mathcal{G}$ a set of initial goals and $Spec = (Pre, Post)$ be a specification respecting constraints and such that $Pre, Post$ are closed under instantiation.

A sufficient condition for $P$ with $\mathcal{G}$ being correct w.r.t. $Spec$ is:

---

[4]  An exact description of the set of allowed calls of constraints is sometimes impossible in our framework, as the set may be not instance closed. For example, many constraints of CHIP have to be called with certain arguments being variables.

1. For each clause $H \leftarrow B_1, \ldots, B_n$ of $P$, $j = 0, \ldots, n$, any substitution $\theta$ and any constraint $c$

$$
\begin{aligned}
&\text{if } \ c \, []\, H\theta \in Pre, \ c \, []\, B_1\theta \in Post, \ \ldots, \ c \, []\, B_j\theta \in Post \\
&\text{then } \quad c \, []\, B_{j+1}\theta \in Pre \quad \text{for } j < n \\
&\qquad\qquad c \, []\, H\theta \in Post \qquad \text{for } j = n
\end{aligned}
$$

2. $\mathcal{G} \subseteq Pre$

*Proof*

Follows from more general Theorem 5.2 applied to a specification set $\{(Pre, Pre \cap Post)\}$. $\quad\square$

For simplicity we consider here only atomic initial goals. Generalization for non atomic ones is not difficult. For instance one may replace a goal $c \, []\, \vec{A}$ by goal $p$ and an additional clause $p \leftarrow c, \vec{A}$ in the program, where $p$ is a new predicate symbol. Alternatively, one can provide a condition for goals similar to that for clauses (Drabent & Małuszyński, 1988), (Apt, 1997).

Notice that the constraints in the clause are treated in the same way as other atomic formulae. As constraint predicates are not defined by program clauses, the requirement that the specification respects constraints is needed in the proposition.

The part of the specification concerning constraint predicates is fixed for a given CLP language. As already mentioned, in our system it is kept in a system library. It is the responsibility of the librarian to assure that the library specification respects constraints. This property depends on the constraint domain in question, and therefore no universal tool can be provided. The number of constraint predicates in any CLP language is finite, so is the library specification, which has only once to be proved to respect constraints.

We want to represent Proposition 3.1 as a system of set constraints. Each implication for a clause $C = H \leftarrow B_1, \ldots, B_n$ from condition 1 of the proposition can now be expressed by a system $F_j(C) = F_{j,1}(C) \cup F_{j,2}(C)$ of constraints, where $F_{j,1}(C)$ consists of

$$
X \ > \ H^{-X}(Call) \ \cap \ \bigcap_{i=1}^{j} B_i^{-X}(Success) \tag{1}
$$

for each variable $X$ occurring in the program clause and $F_{j,2}(C)$ contains one constraint

$$
\begin{aligned}
&Call > B_{j+1} \qquad \text{if } j < n \\
&Success > H \qquad \text{if } j = n
\end{aligned} \tag{2}
$$

(The program variables occurring in the clause become variables of set constraints. As explained in Section 2.2.2, the predicate symbols are treated as function symbols.)

This constraint system has the following property.

*Lemma 3.2*

Let $C = H \leftarrow B_1, \ldots, B_n$ be a clause and $Spec = (Pre, Post)$ a specification. If constraint set $F_j(C)$ has a model assigning to $Call$ the set $Pre$ and to $Success$ the set $Post$ then implication of Proposition 3.1 holds, for any $\theta$ and $c$.

*Proof*
Assume that $I$ is such a model. From (1) it follows that $c \,[\!]\, X\theta \in I(X)$ for each $c, \theta$ satisfying the premise of the implication and for each variable $X$ in the clause. Now from (2) it follows that $c \,[\!]\, B_{j+1}\theta \in I(B_{j+1}) \subseteq Pre$, respectively $c \,[\!]\, H\theta \in I(H) \subseteq Post$ when $j = n$.  □

Set constraints $F_j(C)$ express a sufficient condition for program correctness. If a specification is given, to check the correctness it suffices to check whether the specification extends to a model of $F_j(C)$ (for all $C \in P$ and $j$). In the sequel we show how to do this effectively for the case when $Pre$ and $Post$ are defined by discriminative extended term grammars.

If a specification is not given, Lemma 3.2 tells us that the program is correct with respect to the specification obtained from any model of $F_j(C)$ (for all $C$ and $j$). An algorithm for constructing a discriminative term grammar describing a model of the constraints could thus be seen as a type inference algorithm for this program.

### *3.2  Correctness checking*

In this section we present an algorithm for checking program correctness. We will consider specifications given by means of extended term grammars. Such a grammar $G$ has distinguished variables $Call, Success$ and the specification is $Spec = (\llbracket Call \rrbracket_G, \llbracket Success \rrbracket_G)$ (so $Pre = \llbracket Call \rrbracket_G$, $Post = \llbracket Success \rrbracket_G$). We require that the variables of $G$ are distinct from those occurring in the program. We also require that $Spec$ respects constraints. So such grammar can be seen as consisting of two parts: a fixed part describing the constraints and built-in predicates, and a part provided by the user.

*Example 3.3*
The specification of constraint predicate $\in$ from Example 2.16 can be given by the following grammar rules.

$$
\begin{aligned}
&Call > \in(Any, Nlist) &\quad &Success > \in(Anyfd, Nlist) \\
&Nlist > [\,] &\quad &Anyfd > anyfd \\
&Nlist > cons(Nat, Nlist) &\quad &Nat > nat
\end{aligned}
$$

Consider an atom $B = \in(X, [I, J])$. Applying the generalized projection operation one can compute that $B^{-X}(\llbracket Success \rrbracket) = \llbracket anyfd \rrbracket$ and $B^{-J}(\llbracket Success \rrbracket) = \llbracket nat \rrbracket$.

Notice that within the formalism of extended term grammars we cannot provide a more precise specification. For instance we cannot express the fact that if $c \,[\!]\, \in(t_1, t_2)$ is a success then $c$ constraints the value of $t_1$ to the numbers that occur in the list $t_2$ (formally: any ground element of $cl(\{c \,[\!]\, t_1\})$ is a member of $t_2$).

Our algorithm employs the inclusion check, intersection and generalized projection operations for extended term grammars. As already mentioned, they are rather natural generalizations of the operations for term grammars described in Section 2.1.1. The details can be found in Section 4.4, describing operations for parametric extended term grammars.

The algorithm resembles a single iteration of the iterative algorithm of (Gallagher & de Waal, 1994) for approximating logic program semantics, in its version with "magic transformation". However it works on extended term grammars. We provide its detailed description combined with a proof of its correctness, in order to facilitate a further generalization to parametric case.

As explained in the previous section, a sufficient condition for a program $P$ to be correct w.r.t. *Spec* is that for each $n$-ary clause $C$ of $P$ and for each $j = 0, \ldots, n$, constraints $F_j(C)$ have a model that coincides on *Call* and *Success* with the least model of $G$.

To find such a model we construct (a grammar describing) the least model of $F_{j,1}(C) \cup G$. Then we check if it is a model of $F_{j,2}(C)$. If yes then it is the required model of $F_j(C)$. Otherwise we show that the required model does not exist.

The first step is to compute the projections and intersections of (1). To each expression of the form $A^{-X}(Y)$ occurring in (1) we apply the generalized projection operation to construct a grammar $G_A$ defining $A^{-X}(\llbracket Y \rrbracket_G)$. Then we apply the intersection algorithm to grammars $G_H, G_{B_1}, \ldots, G_{B_j}$. As a result (after appropriate renaming of the variables of the resulted grammar) we obtain a grammar $G_X$ such that

$$\llbracket X \rrbracket_{G_X} = H^{-X}(\llbracket Call \rrbracket_G) \cap \bigcap_{i=1}^{j} B_i^{-X}(\llbracket Success \rrbracket_G).$$

and all the variables of $G_X$, except of $X$, are distinct from those of $F_j(C) \cup G$. Obviously, $\llbracket X \rrbracket_{G_X}$ is the same as $\llbracket X \rrbracket$ in the least model of $\{(1)\} \cup G$.

The first step is to be applied to each constraint (1) of $F_j(C)$ (with a requirement that the variables of the constructed grammars $G_X$ are distinct). Let $G' = \bigcup_X G_X$ be the union of the grammars constructed in the first step. We combine $G'$ and $G$, where the roles of $G', G$ are to define values for, respectively, the variables of $C$ and variables *Call*, *Success*. The least model of $G \cup G'$ is a model of $F_{j,1}(C) \cup G$ (and it coincides with the least model of $F_{j,1}(C) \cup G$ on $Vars(C) \cup \{Call, Success\}$, where $Vars(C)$ is the set of the variables occurring in $C$).

The second step is transforming (2) to a discriminative grammar $G''$, by applying repetitively the construction operation. Let us represent constraint (2) as $Y > A$ (so $Y$ is *Call* or *Success* and $A$ is $B_{j+1}$ or $H$). For each subterm $s$ of $A$, $G''$ employs a variable $X_s$. $X_A$ is $Y$ and if the given subterm $s$ is a variable $V$ then $X_V$ is $V$. Otherwise $X_s$ is a new variable, not occurring in $C, G, G'$. Grammar $G''$ contains the rule $X_s > f(X_{s_1}, \ldots, X_{s_n})$ for each non variable subterm $s = f(s_1, \ldots, s_n)$ of $A$. We have $\llbracket X_s \rrbracket_{G' \cup G''} = \llbracket s \rrbracket_{G'}$, for each subterm $s$. In particular $\llbracket Y \rrbracket_{G' \cup G''} = \llbracket A \rrbracket_{G'} = \llbracket A \rrbracket_{G \cup G'}$.

This completes the construction. We may say that $F_j(C)$ was transformed into a discriminative grammar $F_{C,j} = G' \cup G''$.

It remains to check whether $\llbracket Y \rrbracket_{G' \cup G''} \subseteq \llbracket Y \rrbracket_G$. If yes then $\llbracket A \rrbracket_{G \cup G'} \subseteq \llbracket Y \rrbracket_{G \cup G'}$, i.e. the least model of $G \cup G'$ is a model of $A < Y$. Thus it is the model of $F_j(C) \cup G$ required in Lemma 3.4.

Otherwise, notice first that if $F_1 \subseteq F_2$ then $\llbracket X \rrbracket_{F_1} \subseteq \llbracket X \rrbracket_{F_2}$, for constraint sets $F_1, F_2$. So we have $\llbracket Y \rrbracket_{G' \cup G''} = \llbracket A \rrbracket_{G \cup G'} = \llbracket A \rrbracket_{F_{j,1}(C) \cup G} \subseteq \llbracket A \rrbracket_{F_j(C) \cup G} \subseteq$ $[\![Y]\!]_{F_j(C) \cup G}$. Thus $[\![Y]\!]_{G' \cup G''} \not\subseteq [\![Y]\!]_G$ implies $[\![Y]\!]_{F_j(C) \cup G} \not\subseteq [\![Y]\!]_G$. Hence $I(Y) \not\subseteq [\![Y]\!]_G$ for any model $I$ of $F_j(C) \cup G$ and the required model of $F_j(C) \cup G$ does not exist.

Thus we obtained:

*Lemma 3.4*

The implication from Proposition 3.1 holds for a clause $C$ and a number $j$ if $[\![Y]\!]_{G' \cup G''} \subseteq [\![Y]\!]_G$, for grammars $G', G''$ constructed as above.

The inclusion can be checked by applying the inclusion algorithm (preceded by removing nullable symbols).

We now estimate the complexity of the algorithm. The cost of the intersection operation applied to two grammars with respectively $v_1, v_2$ variables is $O(v_1 v_2)$. The cost of removing nullable symbols is linear (Hopcroft *et al.*, 2001).

Let us now consider the inclusion check. We may assume that grammars are stored so that the productions for each variable are kept together and ordered. Let $v_1, v_2$ be the numbers of variables in the grammars. For each encountered pair $X, Y$ of variables, it has to be checked whether the pair has not occurred previously $(O(\log(v_1 v_2)))$ and the productions for $X$ and for $Y$ are to be found $(O(\log(v_1) + \log(v_2)))$. The pairs of productions with the same function symbol can be found in time proportional to the number of function symbols occurring in the productions found. For each pair of productions $X > f(\ldots), Y > f(\ldots)$ new variable pairs are generated, their number is the arity of $f$. Taking as constants the maximal arity and the maximal number of function symbols in the productions for a given variable, we obtain $O(\log(v_1 v_2))$ per pair. So the total cost of inclusion check is $O(v_1 v_2 \log(v_1 v_2))$. This cost is not changed when the costs of initial sorting of the grammars are taken into account.

Notice that in our algorithm the results of all the generalized projections and intersections computed in the step for $j$ can be reused in the next steps. Taking into account the intersections needed to compute the projections, there are $k - 1$ intersections to be computed for each variable occurring $k$ times in the clause $C$. The cost of computing such a $k$-fold intersection and the size of resulting grammar is $O(v^{k-1})$, where $v$ is the number of variables in the specification grammar $G$.

Computing mappings $\xi$ in the projections and constructing all the $G''$ is linear in the size of the clause. Inclusion checking for a pair of grammars with respectively $O(v^{k-1})$ and $v$ variables can be done in time $O(v^k \log(v^k)) = O(c^k)$, where constant $c$ depends on the number of variables in the grammar.

Thus the correctness checking algorithm described in this section works in time $O(c^k)$, where $k$ is the maximal number of occurrences of a variable in a clause.

*Example 3.5*

Consider the program

```
app([],V,V).
app([A|X],Y,[A|Z]) :- app(X,Y,Z).
```

The verification conditions can be expressed as three constraint systems (we abbre-viate $H = app([A|X], Y, [A|Z])$, $B = app(X, Y, Z)$):

$$V > app([\,], V, V)^{-V}(Call)$$
$$Success > app([\,], V, V)$$

$$A > H^{-A}(Call)$$
$$X > H^{-X}(Call)$$
$$Y > H^{-Y}(Call)$$
$$Z > H^{-Z}(Call)$$
$$Call > app(X, Y, Z)$$

$$A > H^{-A}(Call) \ \cap \ B^{-A}(Success)$$
$$X > H^{-X}(Call) \ \cap \ B^{-X}(Success)$$
$$Y > H^{-Y}(Call) \ \cap \ B^{-Y}(Success) \qquad (3)$$
$$Z > H^{-Z}(Call) \ \cap \ B^{-Z}(Success)$$
$$Success > H$$

Let the following extended term grammar $G$ provide a specification.

$$Call > app(L, L, Any)$$
$$Success > app(L, L, L)$$
$$L > [\,]$$
$$L > [M|L]$$
$$Any > \top$$

where $M$ is further specified by grammar rules not presented here. We assume that $M$ is not nullable in $G$.

Using the described techniques one can check that the specification defines a model for all above stated set constraint systems. For example we check the constraints (3). To compute the projections related to atom $H = app([A|X], Y, [A|Z])$ and *Call* we first obtain the following mapping between the subterm occurrences in $H$ and the variables of $G$.

$$V_{[A|X]} = L \qquad\qquad V_{A^1} = M$$
$$V_{[A|Z]} = Any \qquad\qquad V_{A^2} = Any$$
$$\qquad\qquad\qquad V_X = V_Y = L$$
$$\qquad\qquad\qquad V_Z = Any$$

Similarly, for the projections related to atom $B = app(X, Y, Z)$ and *Success*, we have

$$V_X = V_Y = V_Z = L$$

The grammar describing $H^{-A}(Call)$ is $G \,\dot\cap\, G$ with a distinguished variable $M \,\dot\cap\, Any$. The clauses of $G \,\dot\cap\, G$ for $M \,\dot\cap\, Any$ are $\{\, M \,\dot\cap\, Any \,>\, t \ \mid\ M > t \in G \,\}$. (Also $G \subseteq G \,\dot\cap\, G$.) $M \,\dot\cap\, Any$ is not nullable in $G \,\dot\cap\, G$, as $M$ is not nullable in $G$.

Notice that $B^{-A}(Call) = [\![\top]\!]$ (as $A$ does not occur in $B$). All the other projections from (3) are given by variable $L$ or $Any$ and grammar $G$.

Now we construct grammar $G'$ for which

$$\llbracket A \rrbracket_{G'} = \llbracket M \mathbin{\dot\cap} Any \rrbracket_{G \cap G} \cap \llbracket \top \rrbracket$$
$$\llbracket X \rrbracket_{G'} = \llbracket L \rrbracket_G \cap \llbracket L \rrbracket_G$$
$$\llbracket Y \rrbracket_{G'} = \llbracket L \rrbracket_G \cap \llbracket L \rrbracket_G$$
$$\llbracket Z \rrbracket_{G'} = \llbracket Any \rrbracket_G \cap \llbracket L \rrbracket_G$$

Computing intersections (and renaming variables where necessary) results in a grammar $G'$ consisting of the rules

$$\{\, A > t \mid M > t \in G \,\} \;\cup\; \{\; X > [\,], \qquad Y > [\,], \qquad Z > [\,],$$
$$X > [M|X],\, Y > [M|Y],\, Z > [M|L] \,\}$$

and the rules of $G$ except for those for *Call*, *Success*. (Before constructing the grammar we simplified $\llbracket M \mathbin{\dot\cap} Any \rrbracket_{G \cap G} \cap \llbracket \top \rrbracket$ to $\llbracket M \mathbin{\dot\cap} Any \rrbracket_{G \cap G}$ and $\llbracket L \rrbracket_G \cap \llbracket L \rrbracket_G$ to $\llbracket L \rrbracket_G$. Formally, $G'$ has variables distinct from those of $G$.) Variables $A, X, Y, Z$ are not nullable in $G'$.

The least model of $G'$ provides a valuation for variables $A, X, Y, Z$. It remains to check that for this valuation, together with the valuation for *Success* given by the specification $G$, the constraint $Success > app([A|X], Y, [A|Z])$ holds. To do this we transform this constraint into a discriminative grammar $G''$:

$$Success > app(X_1, Y, X_2)$$
$$X_1 > [A|X]$$
$$X_2 > [A|Z]$$

and apply the set inclusion algorithm to check whether the set defined by *Success* in the specification grammar $G$ is a superset of that defined by *Success* in the obtained grammar $G' \cup G''$. The check succeeds. Hence there exists a model for the considered five constraints which agrees on variables *Call* and *Success* with the model given by the specification. Notice that this holds independently of the missing fragment of $G$ defining $M$.

The same procedure can be performed for all the constraint systems generated for the given program, hence confirming that the program is correct w.r.t. the parametric specification. Also in these cases the correctness check is independent from $\langle M \rangle_G$ (the part of $G$ defining $M$).

In our example the correctness check was independent from a subset $\langle M \rangle_G$ of the specification grammar $G$. This is not uncommon, for some programs and specification grammars a correctness check refers only to some rules of the grammar. Thus a single check is valid for a whole family of grammars. This phenomenon will be exploited in our approach to parametric specifications.

### 3.3 Approximating program semantics

In this work we are mainly interested in checking program correctness. However the representation of the verification condition (Proposition 3.1) as constraints (Lemma 3.2) can be used to obtain an approximation of the semantics of a given program $P$. In the previous section we showed how a single implication from Proposition 3.1 can be expressed by a constraint system $F_j(C)$. We begin with constructing a constraint system representing all the implications from the proposition.

Let us consider the constraints $F_j(C)$ $(j = 1, \ldots, n_C)$ for each clause $C$ of $P$ with $n_C$ body atoms. Let $F_j'(C)$ be $F_j(C)$ with the variables renamed in such a way that the only common variables of (distinct) $F_{j_1}'(C_1)$, $F_{j_2}'(C_2)$ are $Call$ and $Success$. Let grammar $G_0$ specify the initial goals and the constraint predicates. So $[\![Call]\!]_{G_0}$ is the set of initial goals and of the allowed calls of constraints. $[\![Success]\!]_{G_0}$ is (a superset of) the set of possible successes of constraint predicates.[5] Thus $([\![Call]\!]_{G_0}, [\![Success]\!]_{G_0})$ respects constraints.

Now any model $I$ of the constraint system

$$\mathcal{C}(P) = \bigcup_{C \in P} \bigcup_j F_j'(C) \ \cup \ G_0$$

gives a specification $Spec = (I(Call), I(Success))$ with respect to which $P$ is correct, provided that $Spec$ respects constraints. This follows immediately from Lemma 3.2.

In the special case of logic programs a model of $\mathcal{C}(P)$ can be found by using the techniques for set constraint solving. For example the technique of Heintze and Jaffar (1990a; 1991) produces a (non-discriminative) term grammar specifying the least model of set constraints. This technique has been used for generating approximations of logic program semantics (Heintze & Jaffar, 1990b; Heintze, 1992; Heintze & Jaffar, 1994; Charatonik & Podelski, 1998). Another constraint solving approach that uses tree automata techniques, has been presented in (Devienne *et al.*, 1997a; Talbot *et al.*, 2000). We expect that these techniques can be generalized to the case of CLP programs, but we did not investigate this issue yet.

Yet another approach to finding a model of the constraint system $\mathcal{C}(P)$ stems from abstract interpretation techniques (among others (Janssens & Bruynooghe, 1992; Van Hentenryck *et al.*, 1995), (Gallagher & de Waal, 1994), we generalize the latter work in (Drabent & Pietrzak, 1999; Drabent *et al.*, 2000b; Drabent *et al.*, 2000a) and here). $\mathcal{C}(P)$ is seen as a valuation transformer, its fixed points are models of $\mathcal{C}(P)$. Valuations are represented as discriminative grammars. A fixed point is computed iteratively.

To augment our system with a tool for computing approximations of program semantics, we provide a solution based on the latter idea. This choice was guided mainly by possibility of reusing our correctness checking algorithm and the implementation of (Gallagher & de Waal, 1994).

The correctness checking algorithm of the previous section can be easily modified to compute the valuation transformer related to $\mathcal{C}(P)$. This gives an implementation of a single step of the iteration. It remains to combine it with some technique of assuring termination.

**Iteration step.** Take $G_i$ (initially $G_0$). To each $F_j'(C) \cup G_i$ apply the construction as in the correctness checking, obtaining a discriminative grammar $F_{C,j}$. (It is required that all the obtained grammars have distinct variables, except *Call* and *Success*). For each $F_{C,j}$, the variables occurring in $F_{C,j}$ are distinct from those in $G_i$ except for *Call* or *Success*.

---

[5] This approach can also be used when $P$ is a fragment of a program, i.e. the clauses defining some predicates are missing in $P$. Then the semantics of such predicates has to be specified by $G_0$. The algorithm treats them as the constraint predicates. Examples of such program fragments are programs using built-in predicates, unfinished programs or modules of some bigger programs.

The constraints of $F'_j(C)$ are satisfied if the occurrences of *Call*, *Success* in the right hand side of each constraint of the form (1) (Section 3.1) are valuated as in the least model of $G_i$, and the remaining variable occurrences as in the least model of $F_{C,j}$. This follows from the discussion in the previous section.

The obtained grammar $G'_i = G_i \cup \bigcup_{C \in P} \bigcup_j F_{C,j}$ is not discriminative, due to the rules for *Call* and for *Success*. Construct a discriminative approximation of $G'_i$, more precisely a discriminative grammar $G_{i+1}$ such that $[\![Call]\!]_{G'_i} \subseteq [\![Call]\!]_{G_{i+1}}$ and the same for *Success*. This is done by applying the union operation of Section 2.1.1 to $G_i$ and all grammars $F_{C,j}$. (So $G_{i+1}$ is $G_i \dot\cup \bigcup_{C \in P} \dot\bigcup_j F_{C,j}$ with the variable $Call \dot\cup \ldots \dot\cup Call$ renamed into *Call* and $Success \dot\cup \ldots \dot\cup Success$ renamed into *Success*.)

The obtained grammar $G_{i+1}$ has the following property. $\mathcal{C}(P) - G_0$ is true when *Call* and *Success* in all the constraints of the form (1) (Section 3.1) are valuated as in the least model of $G_i$, *Call* and *Success* in the constraints of the form (2) (Section 3.1) as in the least model of $G_{i+1}$, and the (renamed) variables of $P$ as in the least model of $G'_i$.

It remains to check whether the specification given by $G_{i+1}$ does not contain incorrect calls of constraint predicates. This boils down to checking whether all the calls of constraint predicates from the set $[\![Call]\!]_{G_{i+1}}$ are also members of $[\![Call]\!]_{G_0}$. The latter is equivalent to $[\![Call]\!]_F \subseteq [\![Call]\!]_{G_0}$, where $F = G_{i+1} - \{\ Call > A \mid A$ is not a constraint $\}$. Failure of the check means that we are unable to construct a specification which respects constraints. This suggests a program error and an appropriate warning is issued.

This completes an iteration step. Notice that the calls and successes of constraint predicates specified by $G_{i+1}$ are the same as those specified by $G_i$ and thus by $G_0$ (induction on $i$). For calls it follows from succeeding of the checks above. For successes we have that any clause $Success > p(\vec{X})$ from $G_{i+1}$, where $p$ is a constraint predicate, occurs also in $G_i$.

The iteration is terminated if a fixpoint is reached, this means when $[\![Call]\!]_{G_{i+1}} \subseteq [\![Call]\!]_{G_i}$ and $[\![Success]\!]_{G_{i+1}} \subseteq [\![Success]\!]_{G_i}$. (The inclusion in the other direction holds for each $i$). The required model of $\mathcal{C}(P)$ is a valuation in which the values of the variables from $G_0$, except for *Call* and *Success*, are as in the least model of $G_0$, the values of *Call*, *Success* are as in the least model of $G_i$, and the variables of $P$ are valuated by the least model of $G'_i$.

As a result we obtain that whenever the iteration terminates, program $P$ is correct w.r.t. the specification given by the obtained grammar $G_i$.

Notice that this is justified in a different way than usually done in abstract interpretation. Instead of relating a single iteration step to the concrete semantics of the program, we showed that the obtained fixpoint satisfies a sufficient condition for program correctness.

**Termination.** Usually the iterative process described above does not terminate. It should be augmented with means of assuring termination. The idea is to apply a *restriction* operator $\mathcal{R}$ that maps an infinite domain of grammars to its finite subset. Moreover, the operator $\mathcal{R}$ computes an approximation of a grammar $G$ (i.e. $[\![Call]\!]_G \subseteq [\![Call]\!]_{\mathcal{R}(G)}$ and $[\![Success]\!]_G \subseteq [\![Success]\!]_{\mathcal{R}(G)}$). The operator is applied in every iteration step: the newly obtained grammar $G_{i+1}$ is replaced by a grammar $H_{i+1} = \mathcal{R}(G_{i+1})$. In this way we obtain a sequence of grammars $G_0, H_1, H_2, \ldots$, the sequence has the properties described in the previous paragraphs. Since the co-domain of $\mathcal{R}$ is finite, the set of grammars $\{G_0, H_1, H_2, \ldots\}$ is finite and the iteration terminates. This technique can be seen as an instance of widening (Cousot & Cousot, 1992).

An attempt at such approach was made by Gallagher and de Waal (1994). Unfortunately, the termination proof given by the authors is erroneous and Mildner (1999) showed an artificial example which results in an infinite loop.

We adapt a technique presented in (Mildner, 1999), Section 6.5, and inspired by (Janssens & Bruynooghe, 1992). We describe it briefly. Let the *principal label* of a variable $X$ be the set of function symbols occurring in the right hand sides of the rules defining $X$ in a given grammar $G$. Let a *term grammar graph* be a directed graph with grammar variables as vertices. An edge $(X, Y)$ belongs to the graph iff there is a rule $X > f(\ldots, Y, \ldots)$ in the grammar. The operator $\mathcal{R}$ computes an approximation of a grammar $G$ ( $[\![Call]\!]_G \subseteq [\![Call]\!]_{\mathcal{R}(G)}$ and $[\![Success]\!]_G \subseteq [\![Success]\!]_{\mathcal{R}(G)}$) assuring at the same time that there is a spanning tree of the graph of $\mathcal{R}(G)$ such that each branch of the tree contains no more than $k$ variables with the same principal label. Since the grammar is discriminative, and since there is a finite number of function symbols in a program, the set of such spanning trees (modulo variable renaming), is finite and consequently the co-domain of $\mathcal{R}$ (modulo variable renaming) is finite. We usually apply $k = 1$.

The reasoning above does not provide any useful estimation of the complexity of the algorithm. Our experience shows that it is sufficiently efficient to compute directional types of medium size programs.

There exist variants of this method, taking into account a number of occurrences of a single function symbol along a path or just simply binding a depth of the spanning tree with a constant.

Another possibility to cope with the termination problem is to restrict the class of grammars so that the class of defined sets is a partial order of finite heights[6].

---

[6] For example Boye (1996) suggested that the inference is always done with a finite lattice of types. In practice this means that for a class of applications we may have a finite library of types, represented by grammars, which may be extended by need. This will also facilitate communication with the user who will easier understand standard application-specific types than the types represented by automatically generated grammars.

## 4 Parametric Set Constraints

### *4.1 Motivation*

In Example 3.5, the correctness checking of the program was done without referring to a missing fragment $\langle M \rangle_G$ of the grammar that provided the specification. This was due to the fact that the constraints did not include generalized projections of $\langle M, G \rangle$ and all intersections involving $M$ were of the form $M \cap M$ or $M \cap Any$, where *Any* is defined by clause $Any > \top$. The meaning of such expressions is preserved if we simplify them to $M$. As a result we obtained a term grammar referring to $M$. The obtained solution is parametric in the sense that it will hold for any specific choice of the missing fragment of the grammar. Thus the example demonstrates parametric polymorphism of *append*, where calls and successes are approximated by sets determined by the same specific $M$. This kind of parametric polymorphism is useful in locating program errors (cf. the examples in Section 6). In the rest of this section we extend previously introduced basic concepts to be able to handle parameters.

### *4.2 Syntax and Semantics*

To define a notion of a parametric set constraint we extend the alphabet. In addition to the symbols discussed in Section 2.1 we assume that the alphabet also includes *parameters* disjoint with the other categories of symbols. Parameters will be denoted by Greek letters $\alpha, \beta, ....$ A *parametric set expression* is a parameter, a variable, a constant, or it has a form $f(e_1, ..., e_n)$, $t^{-X}(e)$ or $e_1 \cap e_2$, where $f$ is an $n$-ary function symbol, $t$ is a term, $X$ a variable and $e, e_1, ..., e_n$ are parametric set expressions. Notice, that this definition extends the usual definition of set expressions, so that a usual set expression without parameters becomes a special case of a parametric set expression. A parametric term expression is *atomic* if it does not include projection and intersection symbols.

For a given valuation of the variables, a parametric set expression denotes a function from valuations of parameters to subsets of the Herbrand universe. The value of the function for a specific valuation of parameters is determined by considering parameters to be additional variables of the set expression.

We will consider *parametric set constraints* of the form

$$Variable > Parametric \ set \ expression.$$

As discussed above, a collection of non-parametric set constraints has the least model which can be defined by a term grammar. A similar property holds in the parametric case. Take a collection $\mathcal{C}$ of parametric set constraints and treat the parameters as variables. For any given fixed valuation $I$ of the parameters there exists the least model out of the models of $\mathcal{C}$ coinciding with $I$ on the parameters. (This can be proved similarly as Proposition 2.9).

In order to deal with sets of constrained terms parametric set expressions can be generalized to parametric extended set expressions. This is done by permitting base symbols to appear in the expressions. Parametric extended set expressions give rise to parametric extended set constraints. For any fixed valuation of parameters, a collection of such constraints has the least model. (Proof as in Proposition 2.9).

### *4.3 Parametric Term Grammars*

Our parametric specifications will be expressed by parametric grammars. We first introduce parametric term grammars and a notion of an instance of such a grammar. Such instances define sets of terms. Then we extend this approach to define sets of constrained terms.

*Definition 4.1*
A *parametric term grammar* $G$ is a finite collection of parametric set constraints of the form $X > t$ where $X$ is a variable and $t$ is an atomic parametric set expression.

For instance we can consider the grammar $G$ of Example 3.5 as a parametric grammar with one parameter $M$.

In the context of parametric grammars, a (parametric) set descriptor is a pair $\langle X, G \rangle$ where $G$ is a parametric grammar and $X$ a variable or a parameter. The derivability relation is defined in the same way as for non-parametric term grammars. Notice, however, that the normal forms may include parameters.

Parameterless grammars are used to define sets, the role of parametric grammars is to define mappings on sets. This is done by assigning sets to the parameters of a grammar. The sets are given by some other grammars.

Let $G$ be a parametric grammar such that $\alpha_1, \dots, \alpha_k$ are all parameters occurring in $G$. Sometimes we will denote it $G(\vec{\alpha})$ where $\vec{\alpha} = (\alpha_1, \dots, \alpha_k)$. A function $\Phi$ that maps each parameter $\alpha_i$ of $G$ into a set descriptor $\langle X_i, G_i \rangle$ is called, abusing the standard terminology, a *parameter valuation* for $G$. For a given $\vec{\alpha}$ we will sometimes represent a $\Phi = \{ \alpha_1 \mapsto \langle X_1, G_1 \rangle, \dots, \alpha_k \mapsto \langle X_k, G_k \rangle \}$ as the vector $(\langle X_1, G_1 \rangle, \dots, \langle X_k, G_k \rangle)$.

*Definition 4.2*
Let $G$ be a parametric term grammar and let $\Phi = \{\alpha_1 \mapsto \langle X_1, G_1 \rangle, \dots, \alpha_k \mapsto \langle X_k, G_k \rangle)\}$ be a parameter valuation.

An *instance* of $G$ under $\Phi$ is the parametric grammar $G(\Phi) = G' \cup G'_1 \cup \dots \cup G'_k$, where

- $\langle X'_i, G'_i \rangle$ are obtained by renaming apart all variables in each $\langle X_i, G_i \rangle$ so that the grammar $G$ and descriptors $\langle X'_1, G'_1 \rangle, \dots, \langle X'_k, G'_k \rangle$ have pairwise disjoint sets of variables.
- $G'$ is obtained by replacing each parameter $\alpha_i$ in $G$ by $X'_i$.

If $G(\Phi)$ contains no parameters then the usual notion of the sets defined by a grammar applies to $G(\Phi)$.[7] For each its variable $X$ it defines a set, which is $[\![X]\!]_{G(\Phi)}$. So a parametric grammar $G(\alpha_1, \dots, \alpha_k)$ defines a mapping from the sets corresponding to descriptors $\langle X_1, G_1 \rangle, \dots, \langle X_k, G_k \rangle$ to the sets defined by the grammar $G(\Phi)$. Moreover, $G(\Phi)$ defines the value for each parameter $\alpha_i$ of $G$: $[\![\alpha_i]\!]_{G(\Phi)} = [\![X'_i]\!]_{G(\Phi)}$.

---

[7] It applies also to any parametric grammar $H$ and to each variable $X$ such that $\langle X \rangle_H$ is parameterless.

The definition of an instance generalizes in an obvious way from parametric grammars to sets of (extended) parametric set constraints.

*Definition 4.3*
A parametric term grammar is *discriminative* if

- each right hand side of a rule is of the form $f(X_1, \ldots, X_n)$ where each $X_i$ is a variable or a parameter.
- for a given variable $X$ and given $n$-ary ($n \geq 0$) function symbol $f$ there is at most one rule of the form $X > f(\ldots)$

Notice that the instance of a discriminative grammar under a parameter valuation over discriminative grammars is discriminative.

*Example 4.4*
Let grammar $G(\alpha)$ be

$$List > nil \qquad List > cons(\alpha, List)$$

This grammar is discriminative. Consider $\Phi = \{\, \alpha \mapsto \langle List, G \rangle \,\}$. Since $\Phi$ shares variables with $G$ we rename it apart to obtain $\langle List1, G' \rangle$, where $G'$ is:

$$List1 > nil \qquad List1 > cons(\alpha, List1)$$

(The parameters are not renamed, since they are not variables). $G(\Phi)$ is

$$
\begin{aligned}
&List > nil & &List1 > nil \\
&List > cons(List1, List) & &List1 > cons(\alpha, List1)
\end{aligned}
$$

We will use the following notation, when it does not lead to ambiguity. Let $G$ be a discriminative parametric grammar, $X$ a variable and $\vec{\alpha} = (\alpha_1, \ldots, \alpha_k)$ the parameters occurring in $G$. By the *(parametric) type* $X(\vec{\alpha})$ we mean the family of sets defined by $X$ in $G$ (more precisely the mapping from parameter valuations to sets, assigning $[\![X]\!]_{G(\Phi)}$ to $\Phi$). In the special case of a parameterless grammar $G$, type $X$ is the set $[\![X]\!]_G$. Let $\Phi = \{\, \alpha_1 \mapsto \langle X_1, G_1 \rangle, \ldots, \alpha_k \mapsto \langle X_k, G_k \rangle \,\}$ be a parameter valuation, where the grammars are discriminative and the parameters occurring in $G_i$ are $\vec{\alpha_i}$, for $i = 1, \ldots, k$. Then by type $X(X_1(\vec{\alpha_1}), \ldots, X_k(\vec{\alpha_k}))$ we mean the family of sets defined by $X$ in grammar $G(\Phi)$.

For instance the mapping corresponding to variable $List$ in grammar $G(\alpha)$ of the last example can be called $List(\alpha)$. The mapping corresponding to $List$ in $G(\Phi)$ can be called $List(List(\alpha))$.

Instances of parametric discriminative term grammars define sets of terms. Similarly as in the non parametric case, we generalize this formalism to specify sets of constrained terms. Assume a fixed constraint domain $\mathcal{D}$.

*Definition 4.5*
A *discriminative parametric extended term grammar (PED grammar)* $G$ is a finite set of rules of the form

$$X > f(X_1, \ldots, X_n) \qquad \text{or} \qquad X > b$$

where $f$ is an $n$-ary function symbol ($n \geq 0$), $X$ is a variable, $X_1, \ldots, X_n$ are variables or parameters and $b$ is a base symbol. Additionally, for each pair of rules $X > t_1$ and $X > t_2$ in $G$ the sets $[\![t_1^\top]\!]$ and $[\![t_2^\top]\!]$ are disjoint (where $u^\top$ stands for $u$ with each occurrence of a variable or a parameter replaced by $\top$).

The definition of an instance of a grammar applies to parametric extended grammars too. A parameterless instance of such grammar defines a set of constrained atoms for each variable, as described in Section 2.2.2.

*Example 4.6*
Take the grammar $G(\alpha)$ from the previous example. Using $\Phi = \{\alpha \mapsto \langle Any, \{Any > \top\}\rangle\}$ we obtain $G(\Phi)$ defining lists of arbitrary constrained terms. Formally, $\langle List, G(\Phi)\rangle$ defines the set $\{c[\![t_1, \ldots, t_n]\!] \mid n \geq 0, \ t_i \text{ are terms}\}$ (as any term of the form $[\top, \ldots, \top]$ can be generated from $List$ in grammar $G(\Phi)$).

### 4.4 Operations on extended parametric term grammars

We now extend the operations of Section 2.1.2 to extended parametric discriminative term grammars. For each of them we show how the resulting grammar approximates a relevant set operation for each parameterless instance of the arguments.

#### 4.4.1 Emptiness Check and Construction

A variable $X$ in a PED grammar $G$ will be called *nullable* if no variable-free term (i.e. a term consisting entirely of function symbols, base symbols and parameters) can be derived from $X$ in $G$. So for a nullable $X$, $[\![X]\!]_{G(\Phi)} = \emptyset$ independently from $\Phi$. Similarly as in non parametric case, algorithms for finding nullable symbols in context-free grammars can be applied here. Notice that for a non nullable $X$ there exists a $\Phi$ such that $[\![X]\!]_{G(\Phi)} \neq \emptyset$ (provided that the grammar does not contain a base symbol $b$, for which $[\![b]\!] = \emptyset$).

The construction operation extends naturally to parametric grammars. Let $\langle X_1, G_1\rangle, \ldots, \langle X_n, G_n\rangle$ be set descriptors with pairwise disjoint variables and let $f$ be an $n$-ary function symbol. By $f(\langle X_1, G_1\rangle, \ldots, \langle X_n, G_n\rangle)$ we denote set descriptor $\langle Y, G\rangle$, where $Y$ is a new variable and

$$G = \{Y > f(X_1, \ldots, X_n)\} \cup G_1 \cup \ldots \cup G_n$$

(When the set descriptors have some common variables then $f(\langle X_1, G_1\rangle, \ldots, \langle X_n, G_n\rangle)$ can be defined by renaming apart the variables in the descriptors). Clearly:

*Proposition 4.7*
For any parameter valuation $\Phi$ the set descriptors $f(\langle X_1, G_1\rangle, \ldots, \langle X_n, G_n\rangle)(\Phi)$ and $f(\langle X_1, G_1(\Phi)\rangle, \ldots, \langle X_n, G_n(\Phi)\rangle)$ are identical (up to renaming of the variables introduced while building the grammar instances and of the variable introduced by the construction operation).

If $G_1(\Phi), \ldots, G_n(\Phi)$ do not contain parameters then

$$f([\![X_1]\!]_{G_1(\Phi)}, \ldots, [\![X_n]\!]_{G_n(\Phi)}) = [\![Y]\!]_{f(\langle X_1, G_1 \rangle, \ldots, \langle X_n, G_n \rangle)(\Phi)}$$

### 4.4.2 Intersection

Let $G_1$ and $G_2$ be PED grammars. We assume without loss of generality that they have no common variables, but they may have common parameters. We define an operation $\dot{\cap}$ on such grammars; the result is a PED grammar $G_1 \dot{\cap} G_2$. The variables of $G_1 \dot{\cap} G_2$ include the variables of $G_1$, the variables of $G_2$ and new variables corresponding to pairs $(X, Y)$ where $X$ is a variable of $G_1$ and $Y$ is a variable of $G_2$. The latter will be denoted $X \dot{\cap} Y$.

We define $G_1 \dot{\cap} G_2$ to consist of the rules of $G_1$, those of $G_2$ and for each $X > s \in G_1$ and $Y > t \in G_2$ at most one rule as described below.

- $X \dot{\cap} Y > f(s_1 \circ t_1, \ldots, s_n \circ t_n)$ $(n \geq 0)$, provided that $s = f(s_1, \ldots, s_n)$, $t = f(t_1, \ldots, t_n)$ and $s_i \circ t_i$ is the following symbol:

  1. it is the variable $s_i \dot{\cap} t_i$, if $s_i$ and $t_i$ are variables,
  2. it is $s_i$, if $s_i$ and $t_i$ are parameters,
  3. it is the variable $Y$, if one of the terms $s_i, t_i$ is $Y$ and the other is a parameter.

- $X \dot{\cap} Y > u$, provided that at least one of $s, t$ is a base symbol and the following holds. Let us denote $\{s_1, s_2\} = \{s, t\}$ where $s_1$ is a base symbol. Now

  — $s_1 = \top$ and $u = s_2$, or $s_2 = \top$ and $u = s_1$, or
  — $s_2$ is a constant $c \in [\![s_1]\!]$ and $u$ is $c$, or
  — $s_2$ is a base symbol, $[\![s_1]\!] \subseteq [\![s_2]\!]$ and $u = s_1$, or $[\![s_2]\!] \subseteq [\![s_1]\!]$ and $u = s_2$.[8]

Some decisions in this construction are arbitrary. Instead of choosing $s_i \circ t_i$ to be $s_i$ when both $s_i, t_i$ are parameters, one may choose $t_i$. For the case of $s_i, t_i$ being a parameter and a variable one may choose $s_i \circ t_i$ to be the parameter. In the latter case we expect that our choice gives more useful results when further operations are applied to $G_1 \dot{\cap} G_2$, as a variable corresponds to a known set of rules while a parameter does not.

We notice that by construction $G_1 \dot{\cap} G_2$ is a PED grammar and all its parameters (if any) appear in $G_1$ or in $G_2$. The construction guarantees also the following property.

*Proposition 4.8*
For every parameter valuation $\Phi$ such that $G_1(\Phi)$ and $G_2(\Phi)$ are parameterless grammars we have

$$[\![X]\!]_{G_1(\Phi)} \cap [\![Y]\!]_{G_2(\Phi)} \subseteq [\![X \dot{\cap} Y]\!]_{(G_1 \dot{\cap} G_2)(\Phi)}$$

for all variables $X$ in $G_1$ and $Y$ in $G_2$.

---

[8] According to our assumptions on base sets, $[\![s_1]\!] \cap [\![f(\top, \ldots, \top)]\!] = \emptyset$. If $s_2 = f(\ldots)$ then no rule corresponding to $X > s$, $Y > t$ should appear in $G_1 \dot{\cap} G_2$.

*Proof*

Denote $G_1 \dot\cap G_2$ by $G$. It is sufficient to show that if $X \Rightarrow^*_{G_1(\Phi)} t$, $Y \Rightarrow^*_{G_2(\Phi)} u$ and $[\![t]\!] \cap [\![u]\!] \neq \emptyset$ then there exists a term $w$ such that $X \dot\cap Y \Rightarrow^*_{G(\Phi)} w$ and $[\![t]\!] \cap [\![u]\!] \subseteq [\![w]\!]$. The proof is by induction on $\max(|t|, |u|)$ (where $|s|$ is the size of a term $s$).

If $t = \top$ then $X > \top \in G_1$, $X \dot\cap Y \Rightarrow^*_{G(\Phi)} u$ and $u$ is the required $w$. Similarly, $w$ is $t$ in the symmetric case of $u = \top$.

If none of $t, u$ is $\top$ and one of them is a base symbol then the other is a base symbol or a constant. Two cases are possible: $[\![t]\!] \subseteq [\![u]\!]$, rule $X \dot\cap Y > t$ is in $G$ and $w = t$, or $[\![t]\!] \supseteq [\![u]\!]$, $X \dot\cap Y > u \in G$ and $w = u$.

Otherwise $t = f(t_1, \ldots, t_n)$, $u = f(u_1, \ldots, u_n)$ (for some function symbol $f$ of arity $n \geq 0$) and the considered derivations are $X \Rightarrow f(\ldots) \Rightarrow^* t$ and $Y \Rightarrow f(\ldots) \Rightarrow^* u$. Grammar $G$ contains a rule $X \dot\cap Y > f(X_1 \circ Y_1, \ldots, X_n \circ Y_n)$ and $G(\Phi)$ contains $X \dot\cap Y > f(Z_1, \ldots, Z_n)$, where $X_i \circ Y_i = Z_i$ unless $X_i \circ Y_i$ is a parameter. For each $i = 1, \ldots, n$ we have three cases.

1. $X_i \circ Y_i$ is the variable $X_i \dot\cap Y_i$. $X_i \Rightarrow^*_{G_1(\Phi)} t_i$ and $Y_i \Rightarrow^*_{G_2(\Phi)} u_i$. Clearly, $\max(|t_i|, |u_i|) < \max(|t|, |u|)$. By the inductive assumption there exists a term $w_i$ such that $Z_i = X_i \dot\cap Y_i \Rightarrow^*_{G(\Phi)} w_i$ and $[\![t_i]\!] \cap [\![u_i]\!] \subseteq [\![w_i]\!]$.
2. $X_i \circ Y_i$ is a parameter from $G_1$. Then $Z_i \Rightarrow^* t_i$ both in $G_1(\Phi)$ and $G(\Phi)$.
3. $X_i \circ Y_i$ is a variable from $G_1$ or $G_2$. Thus $Z_i \Rightarrow^* t_i$ both in $G_1(\Phi)$ and $G(\Phi)$, or $Z_i \Rightarrow^* u_i$ both in $G_2(\Phi)$ and $G(\Phi)$.

This shows that for $i = 1, \ldots, n$ there exists a $w_i$ such that $Z_i \Rightarrow^*_{G(\Phi)} w_i$ and $[\![t_i]\!] \cap [\![u_i]\!] \subseteq [\![w_i]\!]$. Hence $X \dot\cap Y \Rightarrow^*_{G(\Phi)} f(w_1, \ldots, w_n)$ and $[\![t]\!] \cap [\![u]\!] \subseteq [\![f(w_1, \ldots, w_n)]\!]$. $\square$

*Example 4.9*

Grammar $G_1$ describes parametric non-empty lists and grammar $G_2$ specifies lists of natural numbers:

$$G_1 : \quad NEList > cons(\alpha, List) \qquad G_2 : \quad ListN > nil$$
$$List > nil \qquad\qquad\qquad\qquad ListN > cons(Nat, ListN)$$
$$List > cons(\alpha, List)$$

Computing $NEList \dot\cap ListN$ gives a rule:

$$NEList \dot\cap ListN > cons(Nat, List \dot\cap ListN)$$

The new variable $List \dot\cap ListN$ is defined by the following rules:

$$List \dot\cap ListN > nil$$
$$List \dot\cap ListN > cons(Nat, List \dot\cap ListN)$$

Thus we obtained a non-empty list of natural numbers as a result.

### 4.4.3 Union

Let $G_1$ and $G_2$ be PED grammars. We assume without loss of generality that they have no common variables, but they may have common parameters. We define an operation $\dot\cup$ on such grammars; the result is a PED grammar $G$, denoted $G_1 \dot\cup G_2$. The variables of $G$ include the variables of $G_1$, the variables of $G_2$ and new variables corresponding to pairs $(X, Y)$ where $X$ is a variable of $G_1$ and $Y$ is a variable of $G_2$. The latter will be denoted $X \dot\cup Y$.

Now $G$ consists of the rules of $G_1 \cup G_2$ and, for each $X \dot\cup Y$, of the rules constructed as follows. Let $R = \{\, t \mid X > t \in G_1 \text{ or } Y > t \in G_2 \,\}$. If $\top \in R$ then $G$ contains $X \dot\cup Y > \top$, otherwise:

1. If $f(s_1, \ldots, s_n) \in R$ ($n > 0$) and no other $f(t_1, \ldots, t_n)$ is in $R$ then $G$ contains $X \dot\cup Y > f(s_1, \ldots, s_n)$.

2. For each pair $f(s_1, \ldots, s_n)$, $f(t_1, \ldots, t_n)$ of distinct elements of $R$ ($n > 0$),[9] $G$ contains $X \dot\cup Y > f(s_1 \circ t_1, \ldots, s_n \circ t_n)$, where each $s_i \circ t_i$ is

   - $s_i \dot\cup t_i$, if $s_i, t_i$ are variables,
   - $s_i$, if $s_i = t_i$ and is a parameter,
   - a new variable $V$ otherwise. In this case also rule $V > \top$ is in $G$.

3. $X \dot\cup Y > s$ is in $G$ for each $s \in R$ such that $s$ is a constant or a base symbol and $[\![s]\!] \not\subseteq [\![t]\!]$ for any base symbol $t \in R$, $t \neq s$.

The result of the construction is a PED grammar. Its parameters (if any) may only originate from $G_1$ and $G_2$. The construction is similar to that for discriminative term grammars. The union involving parameters is approximated by $\top$ unless both arguments are the same parameter. This is because we want the construction to approximate the union for all parameter valuations.

*Proposition 4.10*
For every parameter valuation $\Phi$ such that $G_1(\Phi)$ and $G_2(\Phi)$ are parameterless grammars we have

$$[\![X]\!]_{G_1(\Phi)} \cup [\![Y]\!]_{G_2(\Phi)} \subseteq [\![X \dot\cup Y]\!]_{G_1 \dot\cup G_2(\Phi)}$$

for all variables $X$ in $G_1$ and $Y$ in $G_2$.

*Proof*
Denote $[\![X \dot\cup Y]\!]_{G_1 \dot\cup G_2(\Phi)}$ by $R$. It is sufficient to show that if $X \Rightarrow^*_{G_1(\Phi)} s$ or $Y \Rightarrow^*_{G_2(\Phi)} s$, where $s$ is ground, then $[\![s]\!] \subseteq R$. We show this by induction on the derivation length. We can assume that the same renaming of the variables of $\Phi$ has been used in constructing $G_1(\Phi), G_2(\Phi)$ and $(G_1 \dot\cup G_2)(\Phi)$.

Assume that $V \Rightarrow_H s_0 \Rightarrow^*_H s$, where $V = X$, $H = G_1(\Phi)$ or $V = Y$, $H = G_2(\Phi)$. We have two cases.

- $s_0$ is a constant or base symbol (so $s_0 = s$). There is a rule $X \dot\cup Y > s'$ in $G_1 \dot\cup G_2$ such that $[\![s_0]\!] \subseteq [\![s']\!]$. We have $[\![s]\!] \subseteq [\![s']\!] \subseteq R$.
- $s_0 = f(X_1, \ldots, X_n)$ (where $n > 0$), $s = f(u_1, \ldots, u_n)$ and $X_i \Rightarrow^*_H u_i$ for each $i = 1, \ldots, n$. Grammar $G_1 \dot\cup G_2$ contains a rule $X \dot\cup Y > \top$ or $X \dot\cup Y > f(Y_1, \ldots, Y_n)$. In the first case the result is immediate. In the second case the rule have been introduced by clause 1 or clause 2 of the definition of $G_1 \dot\cup G_2$.

---

[9] Notice that for a given $f$ at most two such elements exist.

In the case of clause 1, $(G_1 \dot\cup G_2)(\Phi)$ contains $X \dot\cup Y > f(X_1, \ldots, X_n)$ and $Y_i = X_i$ whenever $Y_i$ is a variable. $X \dot\cup Y \Rightarrow f(X_1, \ldots, X_n) \Rightarrow^* s$ is a derivation of $(G_1 \dot\cup G_2)(\Phi)$, as $H \subseteq (G_1 \dot\cup G_2)(\Phi)$, Hence $[\![s]\!] \subseteq R$.

In the case of clause 2, each $Y_i$ is $s_i \circ t_i$. If $s_i \circ t_i$ is $s_i \dot\cup t_i$ then $s_i, t_i$ are variables, one of them is $X_i$ and by the inductive assumption $[\![u_i]\!] \subseteq [\![s_i \dot\cup t_i]\!]_{(G_1 \dot\cup G_2)(\Phi)}$, as $X_i \Rightarrow^*_H u_i$. If $s_i \circ t_i$ is a parameter then $s_i \dot\cup t_i = s_i = t_i$. In $(G_1 \dot\cup G_2)(\Phi)$ this parameter is replaced by $X_i$. Notice that in this grammar $X_i \Rightarrow^* u_i$. The last possibility is that $s_i \circ t_i$ is a variable $W$ and $W > \top$ is in $G_1 \dot\cup G_2$.

So $(G_1 \dot\cup G_2)(\Phi)$ contains a rule $X \dot\cup Y > f(r_1, \ldots, r_n)$ where $r_i = s_i \circ t_i$ and $[\![u_i]\!] \subseteq [\![r_i]\!]_{(G_1 \dot\cup G_2)(\Phi)}$, for $i = 1, \ldots, n$. Hence $[\![s]\!] = [\![f(u_1, \ldots, u_n)]\!] \subseteq [\![f(r_1, \ldots, r_n)]\!]_{(G_1 \dot\cup G_2)(\Phi)} \subseteq [\![X \dot\cup Y]\!]_{(G_1 \dot\cup G_2)(\Phi)}$.

□

The requirement that $G_1, G_2$ have no common variables is inessential when $G_1 = G_2$. This holds both for $\dot\cap$ and $\dot\cup$ and follows from the proofs of the last two propositions.

*Example 4.11*
Consider the grammars from Example 4.9, $G_1$ specifying parametric non-empty lists and $G_2$ describing lists of natural numbers.

$$G_1 : \quad NEList > cons(\alpha, List) \qquad\qquad G_2 : \quad ListN > nil$$
$$List > nil \qquad\qquad\qquad\qquad\qquad ListN > cons(Nat, ListN)$$
$$List > cons(\alpha, List)$$

The rules defining $NEList \dot\cup ListN$ are

$$NEList \dot\cup ListN > nil$$
$$NEList \dot\cup ListN > cons(V, List \dot\cup ListN)$$
$$V > \top$$

where $V$ is a new variable. There are similar rules for $List \dot\cup ListN$:

$$List \dot\cup ListN > nil$$
$$List \dot\cup ListN > cons(W, List \dot\cup ListN)$$
$$W > \top$$

#### 4.4.4  Generalized projection for parametric sets

Let $\langle Y, G \rangle$ be a set descriptor, where $G$ is a PED grammar, and $t$ be a term. We are going to construct a PED grammar defining (a superset of) $t^{-X}([\![Y]\!]_{G(\Phi)})$.

We first construct a mapping $\xi(t, G, Y)$ assigning to each subterm occurrence $u$ in $t$ a variable or a parameter $V_u$. $V_u$ occurs in $G$ or is a new variable $Any$. Mapping $\xi(t, G, Y)$ has the following properties:

1. $V_t$ is $Y$.
2. If $u = f(u_1, \ldots, u_n)$ $(n \geq 0)$ and $V_u$ is a parameter or $Any$ then $V_{u_1} = \ldots = V_{u_n} = Any$.
3. If $u = f(u_1, \ldots, u_n)$ $(n \geq 0)$ and $V_u$ is a variable of $G$ then

- $V_u > f(V_{u_1}, \ldots, V_{u_n}) \in G$, or
- $V_u > b \in G$, where $b$ is a base symbol, $u \in [\![b]\!]$ and $V_{u_1} = \ldots = V_{u_n} = Any$. (Notice that if $n \neq 0$ then $b = \top$.)

If $\xi(t, G, Y)$ exists then it is unique, because the grammar is discriminative. $\xi(t, G, Y)$ can be constructed by an obvious algorithm similar to that described in Section 2.1.2.

*Proposition 4.12*
Let $G$ be a PED grammar and $G' = G \cup \{Any > \top\}$. Let $t$ be a term and $X^1, \ldots, X^k$ ($k \geq 0$) be the occurrences of a variable $X$ in $t$. If $\xi(t, G, Y)$ exists then

$$t^{-X}([\![Y]\!]_{G(\Phi)}) \subseteq \bigcap_i [\![V_{X^i}]\!]_{G'(\Phi)}$$

for any parameter valuation $\Phi$ such that $G(\Phi)$ is parameterless.

If $\xi(t, G, Y)$ does not exist or $\bigcap_i [\![V_{Z^i}]\!]_{G'(\Phi)} = \emptyset$ for some variable $Z$ of $t$ then $t^{-X}([\![Y]\!]_{G(\Phi)}) = \emptyset$.

*Proof*
Consider a $\Phi$ as above. Let $H = G(\Phi)$ and $H' = G'(\Phi)$.

We begin with showing the following property. Let $c [\!] u$ be a constrained term and $V_u$ be some variable or parameter of $G'$. If $c [\!] u\theta \in [\![V_u]\!]_{H'}$ then $V_u$ satisfies the conditions for $\xi(t, G, Y)$ above (for some $n, V_{u_1}, \ldots, V_{u_n}$).

Assume that $u$ is not a variable (otherwise the conditions hold vacuously) and that $c [\!] u\theta \in [\![V_u]\!]_{H'}$. For $V_u$ being a parameter or $Any$ the conditions trivially hold. Let $V_u$ be a variable of $G$. We have $V_u \Rightarrow^*_{H'} s$ and $c [\!] u\theta \in [\![s]\!]$, where $s = f(s_1, \ldots, s_n)$, $u = f(u_1, \ldots, u_n)$ and $V_u \Rightarrow_{H'} f(X_1, \ldots, X_n)$, or $s$ is a base symbol and $V_u \Rightarrow_{H'} s$. Then a rule $V_u > f(X_1, \ldots, X_n)$, respectively $V_u > s$ exists in $G$; the rule has the required properties.

Now we show that if $c [\!] t\theta \in [\![Y]\!]_H$ then mapping $\xi(t, G, Y)$ exists and for any subterm $u$ of $t$, $c [\!] u\theta \in [\![V_u]\!]_{H'}$. The latter is equivalent to existence of a ground term $s$ such that $V_u \Rightarrow^*_{H'} s$ and $c [\!] u\theta \in [\![s]\!]$.

The proof is by induction. Let $u$ be a subterm of $t$ and

$$U = \{\, u' \mid u \text{ is a proper subterm of } u', \ u' \text{ is a subterm of } t \,\}.$$

Assume that the required mapping exists on $U$. (So $c [\!] u'\theta \in [\![V_{u'}]\!]_{H'}$ for each $u' \in U$ and the conditions for $\xi(t, G, Y)$ are satisfied.) We show that such a mapping exists for $U \cup \{u\}$. It is sufficient to show that $c [\!] u\theta \in [\![V_u]\!]_{H'}$, then it follows that $V_u$ satisfies the conditions for $\xi(t, G, Y)$ from the property discussed above.

If $u = t$ then $c [\!] u\theta \in [\![V_u]\!]_{H'}$ obviously holds. Otherwise there exists a subterm $u' = f(u_1, \ldots, u_n)$ of $t$ such that $u = u_i$ for some $i$, and a ground term $s'$ such that $V_{u'} \Rightarrow^*_{H'} s'$ and $c [\!] u'\theta \in [\![s']\!]$.

If $V_{u'}$ is a parameter or $Any$ then $V_u$ is $Any$ and $c [\!] u\theta \in [\![Any]\!]_{H'}$. The same reasoning is applicable when $V_{u'}$ is a variable of $G$ and $V_{u'} > b \in G$, as then $b = \top$ and $V_u = Any$.

It remains to consider the case of $V_{u'}$ being a variable of $G$ such that $V_{u'} \Rightarrow_{H'}$

$f(V_{u_1}, \ldots, V_{u_n}) \Rightarrow^*_{H'} s' = f(s_1, \ldots, s_n)$. So $V_{u'} > f(V_{u_1}, \ldots, V_{u_n}) \in G$ and $V_{u_i} \Rightarrow^*_{H'} s_i$. From $c \mathbin{[\!]} f(u_1, \ldots, u_n)\theta \in [\![s']\!]$ it follows that $c \mathbin{[\!]} u\theta \in [\![s_i]\!] \subseteq [\![V_u]\!]_{H'}$. This completes the inductive proof.

Thus if $c \mathbin{[\!]} t\theta \in [\![Y]\!]_H$ then $c \mathbin{[\!]} X^i\theta \in [\![V_{X^i}]\!]_{H'}$ for any occurrence $X^i$ of $X$ in $t$. Hence

$$c \mathbin{[\!]} X^i\theta \in \bigcap_i [\![V_{X^i}]\!]_{G'(\Phi)} \quad \text{and thus} \quad t^{-X}([\![Y]\!]_{G(\Phi)}) \subseteq \bigcap_i [\![V_{X^i}]\!]_{G'(\Phi)}.$$

Notice that if $\xi(t, G, Y)$ does not exist or the intersection above is empty then $c \mathbin{[\!]} t\theta \notin [\![Y]\!]_H$ for any $c, \theta$, and $t^{-Z}[\![Y]\!]_H = \emptyset$ for any variable $Z$. □

The proposition suggests the following algorithm to compute a set descriptor $t^{-X}(\langle Y, G \rangle)$ giving an approximation of the set $t^{-X}([\![Y]\!]_{G(\Phi)})$.

1. Compute $\xi(t, G, Y)$.
2. For each variable $Z$ with the occurrences $Z^1, \ldots, Z^k$ in $t$, apply the intersection algorithm for PED grammars (Section 4.4.2) to compute (an approximation of) $\bigcap_i [\![V_{Z^i}]\!]_{G'(\Phi)}$. This results in a grammar $G_Z = G' \mathbin{\hat\cap} \ldots \mathbin{\hat\cap} G'$ and a variable $Z' = Z^1 \mathbin{\hat\cap} \ldots \mathbin{\hat\cap} Z^k$ such that $\bigcap_i [\![V_{Z^i}]\!]_{G'(\Phi)} \subseteq [\![Z']\!]_{G_Z(\Phi)}$
3. If $\xi(t, G, Y)$ does not exist or some $Z'$ is nullable in $G_Z$ then return $t^{-X}(\langle Y, G \rangle) = \langle V, \emptyset \rangle$ as the result (because $t^{-Z}(\langle Y, G \rangle) = \emptyset$, for any $Z$).
4. Otherwise return $t^{-X}(\langle Y, G \rangle) = \langle X', G_X \rangle$

From the last proposition and the appropriate property of the grammar intersection operation it follows that if the algorithm produces $t^{-X}(\langle Y, G \rangle) = \langle V, H \rangle$ then $t^{-X}([\![Y]\!]_{G(\Phi)}) \subseteq [\![V]\!]_{H(\Phi)}$.

### 4.4.5 Inclusion checking for parametric sets

The algorithms for checking inclusion of the sets defined by discriminative term grammars can be generalized to extended parametric grammars.

The problem is stated as follows. Let $G_1$ and $G_2$ be PED grammars. Let $X$ be a variable in $G_1$ and let $Y$ be a variable in $G_2$. We want to check whether $[\![X]\!]_{G_1(\Phi)} \subseteq [\![Y]\!]_{G_2(\Phi)}$ for any valuation $\Phi$ such that $G_1(\Phi), G_2(\Phi)$ are parameterless. We will denote this fact by $\langle X, G_1 \rangle \sqsubseteq \langle Y, G_2 \rangle$ (often abbreviated to $X \sqsubseteq Y$).

We begin with introducing some notions. By $C(X, Y)$ we mean the least set of pairs (of variables or parameters) such that

- $(X, Y) \in C(X, Y)$ and
- if $(X', Y') \in C(X, Y)$, $X' > f(X_1, \ldots, X_n) \in G_1$ and $Y' > f(Y_1, \ldots, Y_n) \in G_2$ then $(X_1, Y_1), \ldots, (X_n, Y_n) \in C(X, Y)$.

An algorithm checking whether $X \sqsubseteq Y$ follows immediately from the following property and from finiteness of $C(X, Y)$.

*Proposition 4.13*
Let $G_1, G_2$ be PED grammars and $X, Y$ be variables of, respectively, $G_1, G_2$. Assume that for each pair $(V, W) \in C(X, Y)$

- if $V$ is a parameter then $V = W$ or rule $W > \top$ is in $G_2$,
- if $V$ is a variable then

  — for each rule $V > f(V_1, \ldots, V_n) \in G_1$ $(n \geq 1)$ there exists a rule $W > f(\ldots) \in G_2$ or $W > \top \in G_2$, and

  — for each rule $V > c \in G_1$, where $c$ is a constant or base symbol, there exists a $W > c' \in G_2$ such that $[\![c]\!] \subseteq [\![c']\!]$.

Then $X \sqsubseteq Y$.

The reverse implication holds provided $G_1$ does not have nullable symbols, $[\![c]\!] \neq \emptyset$ for each base symbol $c$, and if $[\![c]\!] \subseteq [\![W]\!]_{G_2(\Phi)}$ for some $\Phi$, base symbol or constant $c$ and variable $W$ of $G_2$ then $G_2$ contains a rule $W > c'$ where $[\![c]\!] \subseteq [\![c']\!]$. Intuitively, the last condition means that no set $[\![c]\!]$ is described by $G_2$ by more than one rule.

*Proof*

Assume that the conditions are satisfied. For any $(V, W) \in C(X, Y)$ and any derivation $V \Rightarrow^\star_{G_1} t$, where $t$ is a variable-free term, there exists a derivation $W \Rightarrow^\star_{G_2} u$ such that $[\![t]\!]_{G_1(\Phi)} \subseteq [\![u]\!]_{G_2(\Phi)}$ for any $\Phi$. This can be shown by induction on the structure of $t$. If a constrained term $w$ is in $[\![V]\!]_{G_1(\Phi)}$ then $w \in [\![t]\!]_{G_1(\Phi)}$ for some $t$ as above. Hence $w \in [\![W]\!]_{G_2(\Phi)}$, which completes the "if" part of the proof.

Assume that the conditions are not satisfied, for some pair $(V, W) \in C(X, Y)$. We show that for some parameter valuation $\Phi$ there exists a constrained term $t$ such that $t \in [\![V]\!]_{G_1(\Phi)}$ and $t \notin [\![W]\!]_{G_2(\Phi)}$. We enumerate the possible cases, in each of them such $\Phi$ and $t$ obviously exist.

If $V$ is a parameter then $W$ is a different parameter or a variable such that $[\![W]\!]_{G_2(\Phi)} \neq [\![\top]\!]$. For $V$ being a variable we have two cases. $V > f(\ldots) \in G_1$ and no $W > f(\ldots)$ is in $G_2$, or $V > c \in G_1$ and for each $W > c' \in G_2$ $[\![c]\!] \not\subseteq [\![c']\!]$, hence $[\![c]\!] \cap [\![c]\!]' = \emptyset$ (by our restrictions on base sets).

Now it is easy to construct a $u \in [\![X]\!]_{G_1(\Phi)}$ such that $u \notin [\![Y]\!]_{G_2(\Phi)}$ by induction on the definition of $C(X, Y)$ (on the number of applications of the second rule of the definition of $C(X, Y)$ needed to show that $(V, W) \in C(X, Y)$).  $\square$

We illustrate the check by a simple example.

*Example 4.14*

$$
\begin{array}{llll}
G_1: & Y > cons(\alpha, Z) \qquad & G_2: & X > nil \\
     & Z > nil & & X > cons(\alpha, X) \\
     & Z > cons(\alpha, Y) & &
\end{array}
$$

We want to check the inclusion

$$[\![Y]\!]_{G_1(\Phi)} \subseteq [\![X]\!]_{G_2(\Phi)}$$

for arbitrary parameter valuation $\Phi$ such that $G_1(\Phi)$ and $G_2(\Phi)$ are parameterless. For each pair of $C(Y, X)$ the conditions from the proposition are to be checked. $C(Y, X)$ contains $(Y, X), (\alpha, \alpha), (Z, X)$.

Consider $(Y, X)$. For the rule $Y > cons(\alpha, Z) \in G_1$ there exists $X > cons(\alpha, X) \in$

$G_2$. For $(\alpha, \alpha)$ the check is immediate. For $(Z, X)$, the following pairs of rules are found to satisfy the conditions.

$$Z > nil \in G_1, \quad X > nil \in G_2$$
$$Z > cons(\alpha, Y) \in G_1, \quad X > cons(\alpha, X) \in G_2$$

So the check is successfully completed.

### 4.4.6 Set matching

In our approach, a set of allowed calls of a polymorphic procedure will be specified by a set descriptor $\langle Y, G \rangle$ where $G$ is a PED grammar. A particular call $t$ is allowed if there exists a valuation of parameters $\Phi$ such that $t \in [\![Y]\!]_{G(\Phi)}$.

A set of actual calls may be described by another set descriptor $\langle X, H \rangle$, where $H$ is a PED grammar which has no parameter common with $G$.

We want to be sure that all actual calls are allowed. As the specifications are parametric we have to refer to their instances. The question is then, whether for any valuation $\Psi$ of the parameters of $H$ there exists a parameter valuation $\Phi$ for $G$ such that $[\![X]\!]_{H(\Psi)} \subseteq [\![Y]\!]_{G(\Phi)}$. Additionally we are interested in obtaining a possibly small set $[\![Y]\!]_{G(\Phi)}$. We will call this a *set matching* problem.

A solution can be obtained by a modification of the set inclusion algorithm discussed above. In this extension the parameters of $H$ are handled as constants while searching for such bindings of the parameters of $G$ that the inclusion holds.

For a given $X, H$ and $Y, G$ the matching algorithm constructs a parameter valuation $\Phi$ (possibly containing parameters from $H$) such that for any $\Psi$ for which $H(\Psi)$ is parameterless

$$[\![X]\!]_{H(\Psi)} \subseteq [\![Y]\!]_{G(\Phi)(\Psi)}.$$

(This is expressed as $\langle X, H \rangle \sqsubseteq \langle Y, G(\Phi) \rangle$ in the notation of the previous section).

To describe matching we recall how the inclusion algorithm works. Applied to $X$ in $H$ and $Y$ in $G$, it checks the conditions of Proposition 4.13 for each pair $(s, t) \in C(X, Y)$. The difference with the matching algorithm is in the treatment of a $(s, t)$ where $t$ is a parameter (of $G$). In such case the inclusion algorithm answers "no". In matching we want to instantiate the parameters of $G$ so that inclusion holds. So in this case the matching algorithm binds the parameter $t$ to $s$ (which is a variable or a parameter). Notice that several different bindings for $t$ may be produced since $t$ may appear in several pairs in $C$.

As $C(X, Y)$ is finite, the checking terminates with failure or success. In the latter case a set of bindings is produced. From these bindings we now construct a parameter valuation $\Phi$. This is done separately for each parameter $\alpha$. Let $\{\alpha \mapsto s_1, \ldots, \alpha \mapsto s_k\}$ $(k \geq 1)$ be the set of bindings for $\alpha$ produced by the algorithm. The valuation $\Phi(\alpha)$ is constructed by considering the following cases:

- If $k = 1$ then $\Phi(\alpha) = \langle s_1, H \rangle$.
- If $k > 1$ and all $s_i$ are variables of $H$, then $\Phi(\alpha) = \langle s_1 \dot\cup \ldots \dot\cup s_k, H \dot\cup \ldots \dot\cup H \rangle$.
- Otherwise $k > 1$ and some $s_i$ is a parameter. Then $\Phi(\alpha) = \langle X, \{X > \top\} \rangle$ where $X$ is a new variable.

Let $\alpha_1, \ldots, \alpha_n$ be all the parameters of $G$ that appear in $C(X, Y)$. Applying the above stated rules to each of them we obtain $\Phi = \{\alpha_1 \mapsto \Phi(\alpha_1), \ldots, \alpha_n \mapsto \Phi(\alpha_n)\}$.

This completes the description of the matching algorithm. It remains to show that if it succeeds then $[\![X]\!]_{H(\Psi)} \subseteq [\![Y]\!]_{G(\Phi)(\Psi)}$, for any parameter valuation $\Psi$. Assume that $\Phi(\alpha_i) = \langle X_i, G_i \rangle$ (for $i = 1, \ldots, n$) and that $X_i$ was renamed into $X_i'$ while constructing $G(\Phi)$. We apply the inclusion checking algorithm to $X, H$ and $Y, G(\Phi)$ and compare its actions with those of the matching algorithm for $X, H$ and $Y, G$.

Whenever the matching algorithm produces a pair $(s, t)$ of two variables, the same pair is produced by the inclusion checking algorithm. Whenever the former produces an $(s, \alpha_i)$ then the second produces $(s, X_i')$. Grammar $G_i$ has been constructed in such a way that $[\![s]\!]_{H(\Psi)} \subseteq [\![X_i]\!]_{G_i(\Psi)}$. As $[\![X_i]\!]_{G_i(\Psi)} = [\![X_i']\!]_{G(\Phi)(\Psi)}$ we have $[\![s]\!]_{H(\Psi)} \subseteq [\![X_i']\!]_{G(\Phi)(\Psi)}$. Hence for each pair $(s, t)$ produced by the inclusion checking algorithm, $[\![s]\!]_{H(\Psi)} \subseteq [\![t]\!]_{G(\Phi)(\Psi)}$. This completes the proof.

*Example 4.15*
This example illustrates set matching. The parametric grammars $H$ and $G$ specify different variants of lists with elements being triples.

$$
\begin{array}{ll}
H: & L > nil \\
& L > cons(T, L) \\
& T > t(B, N, \gamma) \\
& B > tt \\
& B > f\!f \\
& N > nat
\end{array}
\qquad
\begin{array}{ll}
G: & S > nil \\
& S > cons(E, S) \\
& E > t(\alpha, \alpha, \beta)
\end{array}
$$

We want to match $\langle L, H \rangle$ and $\langle S, G \rangle$. We obtain $C(L, S) = \{(L, S), (T, E), (B, \alpha), (N, \alpha), (\gamma, \beta)\}$. The checks succeed with parameter bindings

$$\{\alpha \mapsto B, \, \alpha \mapsto N, \, \beta \mapsto \gamma\}.$$

The result is the parameter valuation

$$\Phi = \{\alpha \mapsto \langle B \dot{\cup} N, H \dot{\cup} H \rangle, \, \beta \mapsto \langle \gamma, H \rangle\}$$

## 5 Locating Program Errors with Parametric Specifications

The call-success semantics discussed in Section 3 describes a program (together with its set of initial goals) by the set of calls and the set of successes. So the information about which successes correspond to which calls is lost. A more precise semantics can be given by replacing the set of successes by the set of pairs of a call and a corresponding success.

A formalism of distributive grammars does not provide useful approximations of such semantics. If pairs $(call_1, success_1), (call_2, success_2)$ are in such approximation then $(call_1, success_2), (call_2, success_1)$ are there too. Useful approximations can be however provided by parametric distributive grammars. With such a grammar one can specify a family of specifications. Correctness w.r.t. such a family means the following. Whenever a call is correct w.r.t. some specification from the family then any its success is correct w.r.t. this specification. Additionally, each call is correct w.r.t. some of the specifications.

In this section we address the question of partial correctness of programs w.r.t. parametric specifications. First we state the problem and show that it can be re-formulated in terms of parametric set constraints. We show how to employ the constraints to check whether a program is correct w.r.t. a given specification and how to compute a specification for which the program is correct. Then we formalize the notion of error and discuss how the correctness checking procedure locates errors.

### 5.1 Parametric specifications and program correctness

By a parametric specification we mean a set of specifications.[10] We are interested in specifications given by parametric grammars, this is however insignificant for the purposes of this section. Here we define the notion of correctness for such specifications and prove a sufficient condition for such correctness.

*Definition 5.1*
Let $Spec$ be a parametric specification. A call $c \,[\!]\, A$ in an LD-derivation is *correct* w.r.t. $Spec$ if there exists some $(Pre, Post) \in Spec$ such that $c \,[\!]\, A \in Pre$. A success $c' \,[\!]\, A\theta$ corresponding to a call $c \,[\!]\, A$ is *correct* w.r.t. $Spec$ if $c' \,[\!]\, A\theta \in Post$, for any $(Pre, Post) \in Spec$ such that $c \,[\!]\, A \in Pre$.

A program $P$ with a set of initial goals $\mathcal{G}$ is *correct* w.r.t. $Spec$ iff in any LD-derivation of $P$ starting from a goal from $\mathcal{G}$ all the calls and successes are correct w.r.t. $Spec$. A program $P$ is *correct* w.r.t. $Spec$ iff $P$ with the set of initial goals $\bigcup \{\, Pre \mid (Pre, Post) \in Spec \,\}$ is correct w.r.t. $Spec$.

We impose following restrictions on parametric specifications. If $(Pre, Post)$ is a member of such a specification then $Pre, Post$ are closed under instantiation and $Pre \supseteq Post$.[11] The correctness criterion from Proposition 3.1 can now be generalized.

*Theorem 5.2*
Let $P$ be a CLP program, $\mathcal{G}$ a set of atomic initial goals and $Spec$ be a parametric specification. Let each $(Pre, Post) \in Spec$ respect constraints. A sufficient condition for $P$ with $\mathcal{G}$ being correct w.r.t. $Spec$ is:

1. For each clause $H \leftarrow B_1, \ldots, B_n$ and any $(Pre_0, Post_0) \in Spec$ there exist $(Pre_1, Post_1), \ldots, (Pre_n, Post_n) \in Spec$ such that for $j = 0, \ldots, n$, any substitution $\theta$ and constraint $c$

   if $\; c \,[\!]\, H\theta \in Pre_0, \; c \,[\!]\, B_1\theta \in Post_1, \; \ldots, \; c \,[\!]\, B_j\theta \in Post_j$
   then
   $\quad c \,[\!]\, B_{j+1}\theta \in Pre_{j+1}$, if $j < n$
   $\quad c \,[\!]\, H\theta \in Post_0$, if $j = n$

---

[10]  Remember that a (non parametric) specification is a pair of sets of (constrained) atoms.
[11]  The latter condition is not essential. To abandon it, it is sufficient to replace each $Post_l$ in theorem 5.2 by $Pre_l \cap Post_l$.

2. Each element of $\mathcal{G}$ is in some $Pre$, such that $(Pre, Post) \in Spec$.

(As explained in Section 3.1, the restriction to atomic initial goals is not substantial).

*Proof*

Consider the $i$-th goal $Q_i$ of an LD-derivation starting from a goal $Q_0 \in \mathcal{G}$. We show that the call and the successes occurring in $Q_i$ are correct. The proof is by induction on $i$. If $i = 0$ then $Q_i$ contains no successes and the call in $Q_i$ is obviously correct.

Let $i > 0$. Consider the call in $Q_i$. (The case of the goal containing no call is considered later on). $Q_i$ is of the form $c \,[\!]\, (B_{j+1}, \ldots, B_n, \vec{A})\tau$, where $j < n$, for some clause $H \leftarrow B_1, \ldots, B_n$ of $P$, and the derivation is

$$
\begin{array}{rcl}
& & \cdots \\
Q_{i_0} & = & c_0 \,[\!]\, A, \vec{A} \\
Q_{i_1} & = & c_0 \,[\!]\, (B_1, \ldots, B_n, \vec{A})\theta_0 \\
& & \cdots \\
Q_{i_2} & = & c_1 \,[\!]\, (B_2, \ldots, B_n, \vec{A})\theta_0 \theta_1 \\
& & \cdots \\
& & \cdots \\
Q_{i_{j+1}} & = & c_j \,[\!]\, (B_{j+1}, \ldots, B_n, \vec{A})\theta_0 \cdots \theta_j \\
& & \cdots
\end{array}
$$

where $i_{j+1} = i$, $\theta_0 \cdots \theta_j = \tau$ and the call $c_{l-1} \,[\!]\, B_l \theta_0 \cdots \theta_{l-1}$ from a goal $Q_{i_l}$ succeeds in the goal $Q_{i_{l+1}}$ (for $l = 1, \ldots, j$). The calls from $Q_{i_0}, \ldots, Q_{i_j}$ are correct, by the inductive assumption. So there exist $(Pre_0, Post_0), \ldots, (Pre_j, Post_j) \in Spec$ such that $c_0 \,[\!]\, A \in Pre_0$ and $c_{l-1} \,[\!]\, B_l \theta_0 \cdots \theta_{l-1} \in Pre_l$ for $l = 1, \ldots, j$.

Now we show that the successes of these calls are correct. This means $c_l \,[\!]\, B_l \theta_0 \cdots \theta_l \in Post_l$ for $l = 1, \ldots, j$ and for any $(Pre_0, Post_0), \ldots, (Pre_j, Post_j)$ as above. Notice that this includes the $(Pre_1, Post_1), \ldots, (Pre_j, Post_j)$ from condition 1 of the Theorem.

The successes from $Q_{i_2}, \ldots, Q_{i_j}$ are correct by the inductive assumption. Also the success from $Q_{i_{j+1}}$ of $c_{j-1} \,[\!]\, B_j \theta_0 \cdots \theta_{j-1}$ is correct. To show this remove (the instances of) $B_{j+1}, \ldots, B_n, \vec{A}$ from the goals of the derivation $Q_{i_j}, \ldots, Q_{i_{j+1}}$, obtaining a derivation to which the inductive assumption applies. (The derivation is shorter than $i$ and starts from an atomic goal). Other procedure calls (from goals between $Q_{i_j}$ and $Q_{i_{j+1}}$) may succeed in $Q_{i_{j+1}}$. These successes are correct by the same reasoning.

As all $Pre_l, Post_l$ are instance closed, we have $c_j \,[\!]\, A\tau \in Pre_0$ and $c_j \,[\!]\, B_l \tau \in Post_l$ for $l = 1, \ldots, j$. Moreover, $A\tau = H\tau$, as $A\theta_0 = H\theta_0$. From condition 1 of the Theorem it follows that the call $c_j \,[\!]\, B_{j+1} \tau$ is correct.

It remains to consider the case when $Q_i$ does not contain a call. So $Q_i$ is of the form $c \,[\!]\,$ and the initial goal $Q_0$ succeeds in $Q_i$. Let $Q_0 = c_0 \,[\!]\, A$. If $A$ is a constraint then $i = 1$ and $Q_1 = c_0, A \,[\!]\,$. As the specification respects constraints, the success in $Q_1$ is in $Post$ whenever $Q_0 \in Pre$ and $(Pre, Post) \in Spec$. If $A$ is not a constraint then we have a derivation as above, with $j = n \geq 0$ (so $Q_i = Q_{i_{n+1}}$), $Q_{i_0}$ being the initial goal (so $i_0 = 0$) and $\vec{A}$ being empty. Reasoning as previously we obtain that the premises of the implication in the Theorem hold. Hence $c_n \, [\![ \, A\tau = c_n \, [\![ \, H\tau \in Post_0$. As the choice of $Pre_0$ was arbitrary, this holds for any $(Pre_0, Post_0) \in Spec$ such that $c_0 \, [\![ \, A \in Pre_0$. So the success of $c_0 \, [\![ \, A$ is correct. $\quad\square$

In our approach parametric specifications are given by parametric grammars. We assume that such a grammar $G$ has two distinguished variables $Call$, $Success$. The specification is then

$$Spec = \{ \, ([\![Call]\!]_{G(\Phi)}, [\![Success]\!]_{G(\Phi)}) \mid G(\Phi) \text{ is parameterless} \, \}.$$

We require that each specification $(Pre, Post) \in Spec$ respects constraints. Additionally we require that for each $p$ such that $Success > p(\vec{Y}) \in G$, each parameter occurring in $\langle p(\vec{Y}) \rangle_G$ occurs also in $\langle p(\vec{X}) \rangle_G$, where $Call > p(\vec{X}) \in G$. Informally, this means that procedure successes may only depend on those parameters on which the corresponding procedure calls depend. This assures that to each $Pre$ there corresponds exactly one $Post$ such that $(Pre, Post) \in Spec$.

Each grammar providing a specification can be seen as consisting of two parts. One is fixed for a given programming language and specifies the semantics of constraint predicates. The second is given by the user and describes the predicates defined by her program. Real CLP languages have built-in predicates, they can be treated by our method like constraint predicates.

### 5.2  Correctness checking

In this section we discuss checking the verification conditions of Theorem 5.2 with respect to a parametric specification given by a PED grammar. We generalize to such specifications the ideas of Section 3.2.

Similarly as in the parameterless case, each implication from Theorem 5.2 can be expressed by a system $F_j(C)$ of constraints consisting of

$$X \; > \; H^{-X}(Call_0) \, \cap \, \bigcap_{i=1}^{j} B_i^{-X}(Success_i) \tag{4}$$

(where $C = H \leftarrow B_1, \dots, B_n$ is the considered clause, $0 \leq j \leq n$ and $k$ ranges over the occurrences of $X$ in the considered atom) for each variable $X$ occurring in $C$, and of the

$$\begin{array}{ll} Call_{j+1} > B_{j+1} & \text{if } j < n, \\ Success_0 > H & \text{if } j = n. \end{array} \tag{5}$$

So for the condition 1 from the Theorem to hold it is sufficient that for each choice of $(Pre_0, Post_0) \in Spec$ there exist $(Pre_1, Post_1), \dots, (Pre_n, Post_n) \in Spec$ such that each constraint system $F_j(C)$ $(j = 0, \dots, n)$ has a model $I$ in which $I(Call_i) = Pre_i$, $I(Success_i) = Post_i$, for $i = 0, \dots, n$.

Now assume that the specification is given by a parametric grammar $G$. A particular $(Pre_0, Post_0)$ is given by a parameterless instance $G(\Phi)$ of $G$ for some parameter valuation $\Phi$: $Pre_0 = [\![Call]\!]_{G(\Phi)}$, $Post_0 = [\![Success]\!]_{G(\Phi)}$. For any such $\Phi$ we are looking for $\Phi_1, \ldots, \Phi_n$ describing, respectively, $(Pre_1, Post_1), \ldots, (Pre_n, Post_n)$. As the latter depend on $\Phi$, the grammars of $\Phi_1, \ldots, \Phi_n$ may be parametric, with the parameters originating from grammar $\langle Call \rangle_G$. $\Phi_1, \ldots, \Phi_n$ should be chosen in such a way that for any $\Phi$, each $F_j(C)$ has a model $I$ in which

$$
\begin{aligned}
I(Call_0) &= [\![Call]\!]_{G(\Phi)}, & I(Success_0) &= [\![Success]\!]_{G(\Phi)}, \\
I(Call_i) &= [\![Call]\!]_{G(\Phi_i)(\Phi)}, & I(Success_i) &= [\![Success]\!]_{G(\Phi_i)(\Phi)}
\end{aligned}
\tag{6}
$$

for $i = 1, \ldots, n$. This can be done in the following way.

Assume that $\Phi_1, \ldots, \Phi_j$ $(0 \leq j \leq n)$ have already been found. We show how to check the $j$-th implication of Theorem 5.2 and, if $j < n$, how to construct $\Phi_{j+1}$. Let $G_0, \ldots, G_j$ be the grammars $G, G(\Phi_1), \ldots, G(\Phi_j)$ with the variables renamed apart such that

1. $Call$, $Success$ in $G(\Phi_i)$ are renamed into, respectively, $Call_i$, $Success_i$, for $i = 1, \ldots, j$, and $Call$, $Success$ in $G$ into $Call_0$, $Success_0$,

2. no variable occurs in more than one grammar $G_0, \ldots, G_j$ and no variable from clause $C$ occurs in $G_0, \ldots, G_j$.

Now $F_j(C) \cup G_0 \cup \ldots \cup G_j$ is to be converted into a discriminative grammar. For each variable $X$ in the clause, constraint (4) is transformed as described in Section 3.2, by applying generalized projection and intersection operations from Section 4.4.

First for each $A^{-X}(Y)$ occurring in (4), by generalized projection we obtain $\langle X_A, G_A \rangle$ such that $A^{-X}([\![Y]\!]_{(G_0 \cup \ldots \cup G_j)(\Phi)}) \subseteq [\![X_A]\!]_{G_A(\Phi)}$. (Notice that $Y$ is $Call_i$ or $Success_i$, thus $[\![Y]\!]_{(G_0 \cup \ldots \cup G_j)(\Phi)} = [\![Y]\!]_{G_i(\Phi)}$.) Then the intersection operation (followed by appropriate variable renaming) is applied to $\langle X_H, G_H \rangle, \langle X_{B_1}, G_{B_1} \rangle, \ldots, \langle X_{B_j}, G_{B_j} \rangle$, resulting in $\langle X, G_X \rangle$ such that

$$
[\![X]\!]_{G_X(\Phi)} \supseteq H^{-X}([\![Call_0]\!]_{(G_0 \cup \ldots \cup G_j)(\Phi)}) \cap \bigcap_{i=1}^{j} B_i^{-X}([\![Success_i]\!]_{(G_0 \cup \ldots \cup G_j)(\Phi)})
$$

In this way we construct $G_X$ for each variable $X$ of $C$. A renaming is applied so that the variables of the constructed grammars $G_X$ are distinct and $Call_1, \ldots, Call_n, Success_1, \ldots, Success_n$ do not occur in any $G_X$. Let $G' = \bigcup_X G_X$. Notice that $G'$ is discriminative and that, for any $\Phi$, the least model of $(G' \cup G_0 \cup \ldots \cup G_j)(\Phi)$ is a model of $\mathcal{C} = F_j(C) - \{(5)\} \cup (G_0 \cup \ldots \cup G_j)(\Phi)$.

Also, the constraint (5) is converted into a discriminative grammar $G''$ in an obvious way, as described in Section 3.2. Each model of $G''$ is a model of (5), each model of (5) coincides with some model of $G''$ on the variables of (5).

Take an arbitrary $\Phi$ (such that $(G_0 \cup \ldots \cup G_j)(\Phi)$ is parameterless). Let $I_\Phi$ be the least model of $\mathcal{C} = F_j(C) - \{(5)\} \cup (G_0 \cup \ldots \cup G_j)(\Phi)$. We have $I_\Phi(X) \subseteq [\![X]\!]_{G'(\Phi)}$ for any variable $X$ occurring in $C$, and $I_\Phi(Y) = [\![Y]\!]_{G_i(\Phi)}$ for $Y$ being $Call_i$ or $Success_i$, $i = 1, \ldots, j$. Let us represent (5) as $Y > A$, where $Y$ is $Call_{j+1}$ or $Success_0$ and $A$ is, respectively, $B_{j+1}$ or $H$. It holds that $[\![Y]\!]_{G'(\Phi) \cup G''} = [\![Y]\!]_{G'(\Phi) \cup \{(5)\}} = [\![A]\!]_{G'(\Phi)} \supseteq I_\Phi(A)$.

If $j = n$ then $Y$ is $Success_0$, $A$ is $B_{j+1}$ and it remains to apply the inclusion algorithm to check whether

$$\llbracket Success_0 \rrbracket_{G'(\Phi) \cup G''} \subseteq \llbracket Success \rrbracket_{G(\Phi)},$$

for any $\Phi$. If yes then $I_\Phi$ is a model of (5) (as $\llbracket Success \rrbracket_{G(\Phi)} = I_\Phi(Success_0)$), hence a model of $F_j(C)$. It has the required properties, as (6) holds for $i = 1, \ldots, n$.

If $j < n$ then $Y$ is $Call_{j+1}$, $A$ is $B_{j+1}$ and $\Phi_{j+1}$ has to be constructed. We have $I_\Phi(B_{j+1}) \subseteq \llbracket Call_{j+1} \rrbracket_{G'(\Phi) \cup G''}$, for any $\Phi$. Now we apply the set matching operation of Section 4.4 to obtain $\Phi_{j+1}$ such that for any $\Phi$

$$\llbracket Call_{j+1} \rrbracket_{G'(\Phi) \cup G''} \subseteq \llbracket Call \rrbracket_{G(\Phi_{j+1})(\Phi)}.$$

Take an interpretation $I'_\Phi$ such that $I'_\Phi(Call_{j+1}) = \llbracket Call \rrbracket_{G(\Phi_{j+1})(\Phi)}$, $I'_\Phi(Success_{j+1}) = \llbracket Success \rrbracket_{G(\Phi_{j+1})(\Phi)}$ and $I'_\Phi(V) = I_\Phi(V)$ for any other variable $V$. For any $\Phi$, $I'_\Phi$ is a model of (5) (as $I_\Phi(B_{j+1}) \subseteq I'_\Phi(Call_{j+1})$) and hence of $F_j(C) \cup (G_0 \cup \ldots \cup G_j)(\Phi)$. It also fulfills the requirements (6) for $i = 1, \ldots, j$. If the set matching fails, then the program is not found to be correct.

Computing $\Phi_{j+1}$ (or, in the case of $j = n$, performing the inclusion check) completes the iteration step for $j$. The reasoning above provides a proof for:

*Lemma 5.3*

If the process described above succeeds producing $\Phi_1, \ldots, \Phi_n$ then the condition 1. from Theorem 5.2 is satisfied, for clause $C$ and the parametric specification given by the parametric grammar $G$.

If the clause does not satisfy the condition of the Theorem 5.2 then the process of checking is bound to fail. The reverse is not true. The correctness checking of a correct program may fail, due to the fact that the employed intersection and projection operations for parametric grammars are approximate.

Due to similarity of this correctness checking algorithm to that described in Section 3.2, we expect that its complexity is the same.

*Example 5.4*

Consider the following clause, a part of the "Slowsort" program:

```
slowsort(L,S) :- perm(L,S), sorted(S).
```

For this clause we have the following three systems of constraints (we abbreviate *slowsort* as $s$, *perm* as $p$ and *sorted* as $sd$):

$$F_0: \quad L > s(L,S)^{-L}(Call_0)$$
$$S > s(L,S)^{-S}(Call_0)$$
$$Call_1 > p(L,S)$$

$$F_1: \quad L > s(L,S)^{-L}(Call_0) \ \cap \ p(L,S)^{-L}(Success_1)$$
$$S > s(L,S)^{-S}(Call_0) \ \cap \ p(L,S)^{-S}(Success_1)$$
$$Call_2 > sd(S)$$

$$F_2: \quad L > s(L,S)^{-L}(Call_0) \ \cap \ p(L,S)^{-L}(Success_1) \ \cap \ sd(S)^{-L}(Success_2)$$
$$S > s(L,S)^{-S}(Call_0) \ \cap \ p(L,S)^{-S}(Success_1) \ \cap \ sd(S)^{-S}(Success_2)$$
$$Success_0 > s(L,S)$$

A specification is provided by the following parametric grammar $G$:

$$Call > s(ListN, Any) \qquad Success > s(ListN, ListN)$$
$$Call > p(List, Any) \qquad Success > p(List, List)$$
$$Call > sd(ListN) \qquad Success > sd(ListN)$$
$$ListN > [\,] \qquad List > [\,]$$
$$ListN > [Nat|ListN] \qquad List > [\alpha|List]$$
$$Nat > nat \qquad Any > \top$$

The first step of checking the correctness of the clause w.r.t. the specification deals with $F_0$. First one uses generalized projection operation to compute $s(L, S)^{-L}(\langle Call_0, G\rangle) = \langle ListN, G\rangle$ and $s(L, S)^{-S}(\langle Call_0, G\rangle) = \langle Any, G\rangle$. We may informally say that the first two rules of $F_0$ have been transformed into $L > ListN$, $S > Any$.

Then $G_L$ and $G_S$ are respectively $\langle ListN\rangle_G$ and $\langle Any\rangle_G$ (the subsets of $G$ defining $ListN$ and $Any$), with the variables appropriately renamed. Their union is $G'$:

$$L > [\,] \qquad\qquad S > \top$$
$$L > [Nat'|L]$$
$$Nat' > nat$$

The grammar $G''$ is just the last rule of $F_0$. Matching $\langle Call_1, G' \cup G''\rangle \sqsubseteq \langle Call, G\rangle$ succeeds after checking the pairs $(Call_1, Call), (L, List), (S, Any), (Nat', \alpha)$. The result is $\Phi_1 = \{\, \alpha \mapsto \langle Nat', G' \cup G''\rangle \,\}$. So the first implication of the verification condition is satisfied, provided that $(Call_1, Success_1)$ is defined by $G(\Phi_1)$ (after an appropriate variable renaming).

We briefly outline the remaining two steps. Notice that the results of generalized projections from one step are also used in later steps.

Dealing with $F_1$ begins with computing two new generalized projections: $p(L, S)^{-L}(\langle Success_1, G_1\rangle) = \langle List_1, G_1\rangle$ and $p(L, S)^{-S}(\langle Success_1, G_1\rangle) = \langle List_1, G_1\rangle$, where $G_1$ is a renamed $G(\Phi_1)$ and $List_1$ is the renamed $List$. (We may informally say that the first two rules of $F_1$ have been transformed into $L > ListN \cap List_1$, $S > Any \cap List_1$.)

Then intersection operation is applied to approximate sets $[\![ListN]\!]_{G(\Phi)} \cap [\![List_1]\!]_{G_1(\Phi)}$ and $[\![Any]\!]_{G(\Phi)} \cap [\![List_1]\!]_{G_1(\Phi)}$, by grammars $\langle ListN\rangle_G \,\hat\cap\, G_1$ and $\langle Any\rangle_G \,\hat\cap\, G_1$. The grammars are renamed, so that $ListN \,\hat\cap\, List_1$ becomes $L$ and $Any \,\hat\cap\, List_1$ becomes $S$, resulting in $G'$. Matching $\langle Call_2, G' \cup \{Call_2 > sd(S)\}\rangle \sqsubseteq \langle Call, G\rangle$ does not involve any parameter and succeeds, so $\Phi_2 = \emptyset$ and $G_2$ is $G$ with variables renamed.

Similarly, in the third step the projections related to atom $sd(S)$ result in $\langle Any_2, G_2\rangle$ and $\langle ListN_2, G_2\rangle$. (We may informally say that the first rules of $F_2$ have been transformed into $L > ListN \cap List_1 \cap Any_2$, $S > Any \cap List_1 \cap ListN_2$.) Notice that $[\![List_1]\!]_{G_1} = [\![ListN_2]\!]_{G_2}$. $G'$ obtained in this step is essentially the same as that in the previous one – the sets $[\![L]\!]$ and $[\![S]\!]$ that $G'$ defines are the same as in the previous step. The inclusion check succeeds, which completes checking that the clause is correct.

### 5.3 Computing parametric specifications

Now we show how to compute a parametric specification approximating the semantics of a given program.

Consider a parametric specification *Spec*. Notice that if the verification conditions of Proposition 3.1 hold for each (non parametric) specification from *Spec* then the conditions of Theorem 5.2 hold, with $(Pre_0, Post_0) = \ldots = (Pre_n, Post_n)$. Thus the program is correct w.r.t. *Spec*. We will use this fact in constructing parametric specifications for a given program. The initial goals are described by a parametric grammar $G_0$. $G_0$ also describes the constraint predicates, similarly as in Section 3.3. We are going to construct a parametric grammar $G$ (with the parameters from $G_0$) such that whenever the initial call is from $[\![Call]\!]_{G_0(\Phi)}$, all the calls and successes are from $[\![Call]\!]_{G(\Phi)}$, $[\![Success]\!]_{G(\Phi)}$, respectively.

To compute $G$ we proceed as in the parameterless case (Section 3.3). The only difference is that the algorithm is now applied to parametric grammars. We require that the description of constraint predicates is parameterless. So whenever a rule $Call > p(Y_1, \ldots, Y_n)$ or $Success > p(Y_1, \ldots, Y_n)$, where $p$ is a constraint predicate, appears in $G_0$ then $\langle Y_i \rangle_{G_0}$ does not contain any parameters (for $i = 1, \ldots, n$). Obviously, we require that the specification given by $G_0(\Phi)$ respects constraints.

We employ the verification conditions of Proposition 3.1 expressed as the constraint system $\mathcal{C}(P)$ (see Section 3.3). For the grammar $G_0$ as above, $\mathcal{C}(P)$ is parametric. $\mathcal{C}(P) = \mathcal{C}' \cup G_0$, where $\mathcal{C}'$ is a set of parameterless constraints

$$\mathcal{C}' = \bigcup_{C \in P} \bigcup_j F'_j(C).$$

Consider a parameterless instance $G_0(\Phi)$ of $G_0$. If $I$ is a model of $\mathcal{C}(P)(\Phi)$ such that $Spec = (I(Call), I(Success))$ respects constraints then the verification conditions of Proposition 3.1 are satisfied, as shown in Section 3.3.

Our goal is to construct a grammar $G$ such that for any $\Phi$ (for which $G(\Phi)$ is parameterless) there exists a model $I$ of $\mathcal{C}(P)(\Phi)$ in which $I(Call) = [\![Call]\!]_{G(\Phi)}$ and $I(Success) = [\![Success]\!]_{G(\Phi)}$. This implies that the verification conditions of Proposition 3.1 are satisfied for each specification $([\![Call]\!]_{G(\Phi)}, [\![Success]\!]_{G(\Phi)})$. Hence the verification conditions of Theorem 5.2 are satisfied for the parametric specification

$$\{ ([\![Call]\!]_{G(\Phi)}, [\![Success]\!]_{G(\Phi)}) \mid G(\Phi) \text{ is parameterless} \}$$

given by grammar $G$, and the program is correct w.r.t. this specification.

To obtain such a grammar we use the iterative procedure of Section 3.3. It starts with $G_0$ and produces a sequence of grammars $G_i$. Any parameter appearing in $G_i$ occurs in $G_0$. The description of the constraint predicates in any $G_i$ is the same as in $G_0$. The constructed grammars $G_i$ have the following property, for any $\Phi$ (such that $G_0(\Phi)$ is parameterless): The constraints $\mathcal{C}'$ are satisfied if the occurrences of *Call* and *Success* in constraints (1) (see Section 3.1) are valuated as in the least model of $G_i(\Phi)$ and the occurrences of *Call* and *Success* in constraints (2) as in the least model of $G_{i+1}(\Phi)$. This follows from the discussion in Sections 3.2, 3.3, which can be repeated for the case of parametric grammars. The difference is that in the parameter free case the operations of intersection and projection are exact while in the parametric case they are approximate. However the conclusions hold in both cases. In particular, if $G', G''$ are constructed as in Section 3.2 then the least model of $(G \cup G')(\Phi)$ is a model of $F_{j,1(C)} \cup G(\Phi)$ and the least model of $(G' \cup G'')(\Phi)$ is a model of $F_{j,1(C)}$ (for any $\Phi$ assigning parameterless grammars to the parameters of $G$).

As discussed in Section 3.3 it is necessary to apply some technique for enforcing termination while computing fixpoints. As discussed there our prototype implementation uses for that purpose an adaptation of a technique of (Mildner, 1999), which extends also to the parametric case.

Now $G_i$ is the required grammar. For any $\Phi$ as above there exists a model $J$ of $\mathcal{C}'$ which coincides with the least model of $G_i(\Phi)$ on *Call* and *Success*. An interpretation $I$ in which the variables of $\mathcal{C}'$ are valuated as in $J$ and the variables of $G_0$, except of *Call*, *Success*, as in the least model of $G_0(\Phi)$, is the required model of $\mathcal{C}(P)(\Phi)$. As explained above, if such model exists then the program is correct w.r.t. the parametric specification given by $G_i$.

We derive a somehow restricted kind of parametric specifications. Whenever the initial goal is in $[\![Call]\!]_{G(\Phi)}$, all the calls and successes of the computation are, respectively, in $[\![Call]\!]_{G(\Phi)}$, $[\![Success]\!]_{G(\Phi)}$. Thus our approach is unable to construct such parametric specifications that various usages of a predicate in a program are described by different instances of the parametric specification.

### *5.4 Error detection*

The purpose of error diagnosis is to locate the errors in the program. By errors we mean those program fragments that are the reasons that the program is incorrect w.r.t. a given specification. For the semantics chosen in this work, the incorrectness means that some call or success in some computation of the program violates the specification. Such calls or successes will be called error *symptoms*. A pragmatic requirement is that the errors found are as small program fragments as possible.

In traditional approaches, debugging begins with symptoms, obtained from executing the program on some test data. Obviously, only a finite subset of (usually) infinite set of test data can be used. In our approach symptoms are not needed. At the expense of restricting the class of specifications to types defined by parametric discriminative grammars, program correctness can be checked automatically. A successful check is a proof that the program is correct. Equivalently, if the program is incorrect then the check fails; moreover from the correctness checking algorithm we can obtain information locating the errors.

Our correctness checking algorithm uses the sufficient condition of Theorem 5.2. The condition consists of $n + 1$ implications for each $n$-ary clause of the program (and an obvious condition on the initial atomic goals). Each implication concerns a prefix $H \leftarrow B_1, \ldots, B_i$ of a clause $H \leftarrow B_1, \ldots, B_n$ $(1 \leq i \leq n)$.[12] Two implications concern the whole clause $(i = n)$. If the program is incorrect then some of the implications do not hold. The clause prefixes corresponding to these implications will be considered the errors of the program.

---

[12] In the notation of Theorem 5.2, $i = j + 1$ if $j < n$ and $i = n$ if $j = n$.

*Definition 5.5*

Let $P$ be a program and $Spec$ a parametric specification. An *error* in $P$ (w.r.t. $Spec$) is a prefix $H \leftarrow B_1, \ldots, B_{k+1}$ ($0 \leq k \leq n-1$) of a clause $H \leftarrow B_1, \ldots, B_n$ of $P$, or the whole clause $H \leftarrow B_1, \ldots, B_n$ (then $k = n$) such that for some $(Pre_0, Post_0) \in Spec$ and for each $(Pre_1, Post_1), \ldots, (Pre_k, Post_k) \in Spec$ such that the implication of Theorem 5.2 holds[13] for $j = 0, \ldots, k-1$, there exists a substitution $\theta$ and constraint $c$ such that $c \,[\!]\, H\theta \in Pre_0$, $c \,[\!]\, B_1\theta \in Post_1$, $\ldots$, $c \,[\!]\, B_k\theta \in Post_k$ and

$$c \,[\!]\, B_{k+1}\theta \notin Pre_{k+1} \text{ for any } (Pre_{k+1}, Post_{k+1}) \in Spec, \text{ if } k < n,$$
$$c \,[\!]\, H\theta \notin Post_0, \text{ if } k = n.$$

We say that the *representative* of the error is $B_{k+1}$ when $0 \leq k \leq n-1$, or $H$ when $k = n$. (So it is the atom whose instance is found incompatible with the specification).

This definition formalizes the intuition of a program fragment being the reason of incorrectness. Such fragments have to be changed in order to obtain a correct program. On the other hand, in a general case there are no semantic criteria to state what in such a fragment has to be changed. In this sense the errors defined above are minimal. What is "the error" from the pragmatic point of view, depends on the programmer's intentions about the exact intended semantics of the program.

*Example 5.6*

Consider a type specification

$$Call > m(Any, L) \qquad L > [\,] \qquad\qquad Any > \top$$
$$Success > m(\alpha, L) \qquad L > [\alpha | L]$$

and a clause `m( X, [Y,Z] ) :- m( X, Z )`. The (prefix being the) whole clause is incorrect w.r.t. the specification, as for $j = 0$ the second argument of the call $m(X, Z)\theta$ is, speaking informally, of type $\alpha$ instead of $L$. We cannot state which atom of the clause is erroneous. To obtain a correct clause one may for instance replace $m(X, [Y, Z])$ by $m(X, [Y | Z])$, or $m(X, Z)$ by $m(X, [Z])$. Only knowing that $m$ is intended to define a list membership relation, makes it possible to decide what is the actual error (w.r.t. the (exact) intended semantics of the program).

Notice that there is at most one error in a given clause, as Definition 5.5 requires that the implications for $j = 0, \ldots, k-1$ hold. Thus according to our definition each proper prefix of an error is not an error. The reason is that if $H \leftarrow B_1, \ldots, B_{j+1}$, $0 \leq j < k$, were an error then we would not have a criterion which $(Pre_{j+1}, Post_{j+1})$ to consider in determining that $H \leftarrow B_1, \ldots, B_{k+1}$ is an error.[14]

We will use the correctness checking procedure from the previous section to locate errors in programs. If a clause contains an error then the procedure will fail. The reverse is not true, correctness checking of a clause not containing an error may fail, due to approximation inaccuracies of the intersection and projection operations.

---

[13] This means that for any substitution $\theta$ and constraint $c$
      if   $c \,[\!]\, H\theta \in Pre_0$, $c \,[\!]\, B_1\theta \in Post_1$, $\ldots$, $c \,[\!]\, B_j\theta \in Post_j$
      then   $c \,[\!]\, B_{j+1}\theta \in Pre_{j+1}$
[14] Such a criterion may be obtained by setting $Post_{j+1} = \bigcup \{ Post \mid (Pre, Post) \in Spec \}$.

The correctness checking procedure finds each clause containing an error. Moreover, to a certain extent a clause prefix containing the error is located. If $\Phi_1, \ldots, \Phi_j$ are successfully constructed then each prefix $H \leftarrow B_1, \ldots, B_i$, for $i = 1, \ldots, j$ is not an error. If then constructing of $\Phi_{j+1}$ fails, it is possible that some of prefixes $H \leftarrow B_1, \ldots, B_i$, where $i > j$, is an error. If no approximation inaccuracies had appeared then $H \leftarrow B_1, \ldots, B_{j+1}$ would have been an error. The inaccuracies make it possible that some larger prefix is an error or the clause does not contain an error.

## 6  The prototype diagnosis tool

### *6.1  The structure of the tool*

We implemented a prototype tool that locates errors by checking correctness of a program wrt types specified by PED grammars. Notice that such a grammar may or may not include parameters. As already mentioned in the Introduction, the tool consists of three main components:

- *the type inferencer* – for a given program and parametric entry declaration constructs parametric directional types of the program using the technique of Section 5.3. The types approximate the program semantics.
- *the type checker* – checks correctness of a program wrt to given parametric directional types using the technique of Section 5.2
- *the specification editor* – a GUI which makes it possible to specify intended directional types and also to inspect and to re-use in this specification the inferred types.

A diagnosis session starts with type inference. The inferencer may issue some warnings about illegal calls to built in predicates. It happens if the inferred call type for a built-in is not a subtype of the expected one. The expected call types for built-ins are stored in the system library and may be viewed as a part of specification given a priori.

The main part of the session consists in providing/editing by the user a specification of the intended types. The type checker works interactively with the editor. Each verification condition is checked as soon as a sufficient fragment of a specification is provided. The diagnosis relies entirely on the provided types. It does not involve execution of the program and it does not use the inferred types. The role of type inference is auxiliary. As mentioned above, the inferencer may discover certain irregularities in the program and its warnings suggest starting points for the diagnosis. On the other hand, the inferred types may be used as a draft for the specification; this simplifies the task of constructing the specification by the user.

---

We expect however that the definition modified in such way would define errors which do not correspond to an intuitive notion of an error.

The current version of the tool supports a substantial subset of the CHIP language. It can be easily modified to be used with any Prolog-like language. The prototype has been implemented in SICStus Prolog. A more detailed description of our tool, in its version for parameterless specifications, together with an example error diagnosis session is given in (Drabent *et al.*, 2000a).

### *6.2 Types*

The parametric specifications used by the tool are PED-grammars defined in Section 4.3. For every parameter valuation such a grammar defines a set of constrained terms. A parametric type defined by such a grammar can be seen as a family of sets (of constrained terms).

In the implementation we use the notation as shown in the example below. We write

```
:-typedef tree --> nil; t(elem,tree,tree)
```
to denote the grammar

$$Tree > nil$$
$$Tree > t(Elem, Tree, Tree)$$

Such a grammar may be a part of a program.

The present version of the tool uses four *base* types:

- `any` denotes $[\![\top]\!]$,
- `nat` denotes the set of natural numbers,
- `anyfd` denotes the set of constrained atoms of the form $x \in \text{FD} [\![ ]\!] x$ where FD is a finite domain, i.e. a finite set of natural numbers[15],
- `int` denotes the set of integers.

The approach to base types in the implementation does not satisfy Requirement 2.13. Namely, sets denoted by `anyfd` and `int` are neither disjoint nor one of them includes the other. This design choice remained from the previous versions of our approach. It is dealt with by some ad hoc modifications of the grammar operations. It will be changed, by adding a base type *neg* of negative numbers and defining the set of integers as the union of $[\![nat]\!]$ and $[\![neg]\!]$.

The type of a top call for a program is provided with `entry` declaration, for instance:

```
:- entry delete(list(A),A,any).
```

Parameters are identifiers written with capital letter (like variables in Prolog). Thus the above declaration says that we intend to delete an element of an arbitrary type `A` (the second argument) from the list of elements of that type (the first argument). The third argument is supposed to be a variable on call, which can be only expressed as `any`.

To make the system interface more user-friendly we introduced a library of type definitions which may be augmented by the user. It contains for instance, a parametric grammar defining type `list(A)`, i.e. lists of elements of type `A`.

---

[15] We do not distinguish between $c$ and $x \in \{c\} [\![ ]\!] x$.

Whenever possible, the types computed by the system are presented to the user in terms of those defined in the library or declared by the user. In this way the user faces familiar and meaningful type names instead of artificial ones. For instance, assume that the system has to display a type `t77` together with the grammar rule `t77 --> [];[t78|t77]`. Then it finds that they are an instance of the rules defining `list(A)` and displays `list(t78)` instead.

When providing the specification the user gives intended call and success types for a given program. Formally this means providing grammar rules for *Call* and *Success*. So the grammar providing the specification consists of the rules kept in the library, the grammar rules given in `:-typedef` declarations of the program and the rules for *Call* and *Success* provided by the user during the diagnosis section.

### 6.3 Inferring and checking types

The type inference algorithm is based on the description of Sections 3.3 and 5.3. It computes an approximation of call-success semantics of a given program. This is done by means of fixed point iteration. The algorithm is implemented in Prolog.

For all programs used in our experiments (up to 230 clauses and 52 predicates) the prototype implementation computes approximations in reasonable time[16].

As already mentioned in Section 3.3, in the parameterless case the algorithm can be seen a method of solving set constraints. However, the solution obtained is in general not the least one because of widening and of the approximate nature of the union operation which is used by the algorithm. Extension to the parametric case introduces additional loss of information caused by the operations on PED grammars discussed in Section 4.4.

The type inferencer is not able to find polymorphic dependencies between variables by itself. The only parameters that may appear during the analysis are those provided by the user in the entry declaration.

As discussed in Section 4.4, the definitions of operations on PED grammars include some arbitrary decisions. The union and the intersection of a type parameter with another type are, respectively, $[\![\top]\!]$ and the other type. The implementation produces a warning whenever these situations appear during type inference.

The rationale behind the warnings is as follows. The type parameter in call specification reflects the intuition that any instance of the parametric type is allowed at call. Normally it means that the analyzed procedure is polymorphic and it is supposed to work for any instance of the parameter. Thus the result of the analysis should be independent on potential instantiations of the parameter. In other words, none of the operation on types should touch parameters. If it happens then the procedure may not work as a polymorphic one.

The type inference algorithm constructs call and success types of the predicates defined by program clauses, thus computing an approximation of their call-success semantics. To be able to deal with real programs, it uses a library of type specifications of built-in predicates. Similarly it is able to deal with fragments of programs (for instance with programs under development). In the latter case the user is required to provide type descriptions for the undefined predicates.

---

[16] 21.88 s in the worst case, running SICStus Prolog, ver.3.8.4 on Sun-Ultra 10/440, with 440 MHz CPU speed and 265 MB RAM.

As already mentioned, the diagnosis relies on the type specification provided incrementally by the user. The specification process is supported by the possibility to accept some types constructed in the analysis phase as specified ones. This possibility is restricted to the types of the predicates relevant for the diagnosed predicate. Moreover, a heuristics is used to suggest to the user the order of specifying types. Following this order often results in fewer type specifications needed to locate an error. The user may stop the diagnosis with the first error message, which is often obtained without specifying all requested types. The diagnosis process may be continued by specifying all requested types. In this case, the tool will locate all incorrect clause prefixes in the fragment of the program relevant for the diagnosed predicate.

An error message contains an incorrect clause. The incorrect prefix is indicated by referring to its representative (cf. Definition 5.5). The specification provided by the user is stored by the diagnoser and may be re-used during further diagnosis sessions.

### *6.4 Examples*

Below we show some examples illustrating the use of the diagnosis tool. The examples exhibit an advantage of parametric analysis over the non-parametric one.

Consider the following erroneous program:

```
append([],Ys,Ys).
append([H|Xs],Ys,[H,Zs]) :-
    append(Xs,Ys,Zs).
```

The head of the second clause should be `append([H|Xs],Ys,[H|Zs])`. Assume that the `append/3` predicate is supposed to concatenate two lists of any arbitrary type. In the non-parametric framework the best way to express such a type is `list(any)`. After analyzing the program with the following entry point declaration:

```
:-entry append(list(any),list(any),any).
```

the inferred success type is

```
append(list(any),list(any),list(any))
```

The reason for inferring such a (success) type for the third argument of `append/3` is that the type of two-element list originating from the head of the second clause (`[H,Zs]`) has been joined, by means of the upper bound operation, with the type `list(any)` coming from the recursive call of `append/3`. It results in the type `list(any)`. Thus nothing suspicious can be concluded.

On the other hand, if we provide a parametric declaration:

```
:-entry append(list(A),list(A),any).
```

then the inferred success type does not meet our expectations:

```
append(list(A),list(A),list(any))
```
as we would rather wish to have `list(A)` as a result. Moreover, the analyzer warns us that the parameter `A` (originating from the success of the first clause and type `list(A)` of `Ys`) will be approximated by `any` while computing an upper bound with the type `list(A)` (originating from the second clause and the term `[H,Zs]`, in which `Zs` is of type `list(A)`.

After the user has specified the success type, the diagnoser locates the error and reports it by indicating its representative `append([H|Xs],Ys,[H,Zs])`.

The next example is a fragment of a job scheduling program. The fragment sets up precedence constraints among the jobs. A job is described by a term `job(T,P)`, where `T` is a starting time of processing the job and `P` is its duration. As `T` has to be found by the program it is a domain variable; `P` is fixed. The jobs are kept in a list and are identified by the position in it.

The precedence between two jobs is represented as a term `prec(J1,J2)`, with a meaning: `J2` cannot start before `J1` has been completed. All such pairs are kept in the list. The precedence constraints are set up by the procedure `precedences/2` defined below.

```
:-typedef tprec --> prec(nat,nat).
:-typedef tjob --> job(anyfd,nat).

:-entry precedences(list(tprec),list(tjob)).

precedences([],_).
precedences([prec(A,B)|Ps],Jobs) :-
    get_nth(Jobs,A,job(TA,PA)),
    get_nth(Jobs,B,job(TB,_)),
    TB #>= TA + PA,
    precedences(Ps,Jobs).

get_nth([_|X],1,X) :-!. % bug here
get_nth([_|Xs],N,X) :-
    N1 is N - 1,
    get_nth(Xs,N1,X).
```

The `:-typedef` declaration defines new types used in the entry declaration. The first clause defining `get_nth/3` contains a bug, as the first argument of its head should be `[X|_]`.

The inferred success type for `precedences/2` is:
```
precedences(t52,list(tjob))
```
together with a definition of `t52`:
```
t52-->[]
```
This means that the procedure may succeed only when the precedence list is empty. If a diagnosis session is started with this predicate the user is asked to provide expected call and success types for `get_nth/3`. Assume they are respectively:

```
get_nth(list(A),int,any)
```
and
```
get_nth(list(A),int,A)
```
After this step the diagnoser presents as an error the clause prefix pointed by the representative
```
get_nth([_|X],1,X).
```

The reason for the error message is that inclusion check of `list(A)` and `A` fails. Notice however, that in non-parametric framework the specification for `get_nth/3` could be `get_nth(list(any),int,any)`, both for calls and successes. In this case the inclusion check of `list(any)` and `any` would succeed, and the bug would not be discovered by the diagnosis.

## 7 Discussion and Conclusions

### *7.1 Related Work*

This work is directly related to:

- the research on proving partial correctness of logic programs wrt call-success specifications,
- the research on approximating semantics of logic programs by descriptive types based on set constraints and on abstract interpretation.

It extends some of the techniques proposed in these fields to handle parametric polymorphism and constraint domains.

**Partial correctness.** From (Bronsard *et al.*, 1992; Apt, 1993; Bossi & Cocco, 1989) and our own previous work (Drabent & Małuszyński, 1988; Boye & Małuszyński, 1997) we extend to CLP a *directional* view of logic programs in the sense that each predicate is considered a procedure which, when applied to a suitable tuple of call arguments returns upon a success a tuple of computed values. This is formalized by the notion of call-success semantics.

We rely on the proof methods of (Drabent & Małuszyński, 1988; Bossi & Cocco, 1989) for proving partial correctness of logic programs wrt call-success specification. We use their modification for CLP described in (Drabent *et al.*, 2000b; Drabent *et al.*, 2000a) and we extend them to deal with parametric specifications. For specifications formulated as *definite set constraints* (Heintze & Jaffar, 1990a)[17] correctness can be effectively checked by reformulation of the verification conditions of the above mentioned methods, also as definite set constraints. As discussed in Section 3.1 such a reformulation requires specific operation called *generalized projection*, which is a special case of the "quantified set expression" of (Heintze & Jaffar, 1994) and "membership expression" of (Devienne *et al.*, 1997b; Talbot *et al.*, 2000). For the reasons discussed in Section 2.1.1 we choose as our specification language a parametric variant of well-known formalism of discriminative regular term grammars[18] see e.g. (Dart & Zobel, 1992) additionally equipped with basic types for handling constrained terms and atoms of CLP. The same language is used for describing approximations of call-success semantics. Traditionally such approximations are called *descriptive types* of logic programs.

---

[17] Later studied also by (Charatonik & Podelski, 1997) and (Talbot *et al.*, 2000).

Soundness of our method of type checking is stated by Lemma 3.2. which gives a sufficient condition for correctness of CLP programs for specifications given as term grammars. This result extends then for PED grammars. A recent paper (Comini *et al.*, 2000) argues that such sufficient conditions for verification of Logic Programs can be systematically derived if the considered class of specifications is defined as an abstract interpretation domain with Galois connection relating them to a concrete semantics of logic programs. Unfortunately, as shown in (Drabent & Pietrzak, 1998), for our non-parametric specifications such a Galois connection does not exist,[19] so that it is not clear whether the method is applicable.

**Types in logic programming.** We follow the *descriptive typing* approach where types approximate a posteriori the semantics of untyped programs. The early work on descriptive types (Mishra, 1984; Janssens & Bruynooghe, 1992; Frühwirth *et al.*, 1991; Yardeni & Shapiro, 1991) was based on the least model semantics. The problems considered were how to check that the least model semantics is included in a regular set of terms (the type checking problem) and how to approximate it by regular sets (the type inference problem). The regular sets were defined by regular grammars or equivalently by regular unary logic programs (Frühwirth *et al.*, 1991). This approach does not take into account the intended use of the predicates and gives therefore a few possibilities for finding typing errors. The focus is mostly on detecting that for some predicates the inferred types are empty sets in which case the predicates never succeed.

Checking of directional types based on set constraints was discussed in (Aiken & Lakshman, 1994). The types used are sets of non-ground terms. They are specified by set constraints together with a lifting function *Sat* that maps a set of ground terms to a set of nonground terms. Type checking is based on the same verification condition we use, which in general form originates from (Drabent & Małuszyński, 1988; Bossi & Cocco, 1989) and was specifically formulated for directional type checking in (Apt, 1993). We also allow nonground types but in contrast to this work we achieve non-groundness not by lifting ground sets but by extending set constraints with constants interpreted as basic nonground types.

Inference of directional types in the framework of set constraints was illustrated by an example in (Heintze, 1992). (The main topic of the paper are implementation techniques for solving set constraints.) In the example the types are inferred by constructing set constraints analogous to our encoding of verification conditions, and solving them. A more recent work on inference of directional types for logic programs is (Charatonik & Podelski, 1998). It rephrases (as Theorem 1) the verification conditions of (Apt, 1993) in model-theoretic setting[20]. Thus, a starting point for type inference are again the verification conditions used both in (Aiken & Lakshman, 1994) and in our work. In (Charatonik & Podelski, 1998) the directional types are regular, but in general not discriminative. They are characterized by the least model of a *uniform program* constructed from the original program. The authors are not specific about the algorithms to be used for constructing a representation of the resulting directional types. In contrast to this work we do not construct uniform programs. We encode the verification conditions as set expressions. Directional types are models of these expressions. We restricted our attention to discriminative directional types. This made it possible to extend the type checking and type inference algorithms of (Gallagher & de Waal, 1994; Mildner, 1999), based on abstract interpretation, to the case of parametric directional types.

---

[18] Such grammars define sets acceptable by deterministic root-to-frontier tree automata. Alternatively, the sets are called tuple-distributive or path-closed.

[19] The abstraction function does not exist, as there does not exist the best approximation of a given set of terms by a regular set of terms. This holds for both kinds of regular sets, those defined by discriminative and by arbitrary term grammars.

Our work follows the idea of using semantic approximations for program verification and for locating errors presented in (Bueno *et al.*, 1997). This idea was also used for designing a generic preprocessor for validation and debugging of CLP programs (Puebla *et al.*, 2000). The preprocessor verifies various assertions, provided by the user or inferred, in particular also non-parametric discriminative directional types similar to ours.

While most of the papers on types in logic programming claim error detection as their objective, a little attention is usually devoted to locating errors. In this paper we extend our previous approach to locating errors (Drabent *et al.*, 2000b; Drabent *et al.*, 2000a) to the case of polymorphic types. As discussed in Section 6 this gives some more opportunities to locate the reasons of discrepancy between actual program and user expectations.

**Parametric Polymorphism in Logic Programming.** Use of parametric polymorphic types in logic programming was first suggested in (Mycroft & O'Keefe, 1984). In this approach the function symbols and the predicates of a logic program are supposed to have a priori declared types. The types are used to restrict the syntax of the language to *well-typed* formulae. A compile-time test is then formulated which gives a sufficient condition that well-typedness is an invariant of goals in all computations. This approach to using types, called *prescriptive* typing has been followed in many papers and in several logic programming languages, most notably Gödel (Hill & Lloyd, 1994) and Mercury (Somogyi *et al.*, 1996). Semantically, prescriptive typing corresponds to taking many sorted typed logic as a foundation of logic programming, instead of untyped logic. Our approach is based on untyped logic and our parametric types approximate actual or intended semantics of the program. Thus, our work is in the framework of *descriptive* types, and the vast literature on prescriptive types is not further discussed here. Let us only mention some recent research on this topic (Fages & Coquery, 2001; Smaus *et al.*, 2000; Deransart & Smaus, 2001).

---

[20] These conditions are stated as *magic transformation* of the original program.

In the context of descriptive typing some preliminary ideas on the issue of parametric polymorphism are discussed already in (Mishra, 1984) as a possible extension of the presented type checking method for non-directional types. (Zobel, 1987) presents a method for deriving "syntactic" polymorphic types. These types are not directional. They are clearly related to term grammars but the paper does not explain the relationship. Our techniques focus on directional types and are based on semantic considerations.

Polymorphic directional types for logic programs discussed in (Boye, 1996) are based on the *annotation method* of (Deransart, 1993) for proving correctness of logic programs. This method is different from that used in our work and refers to a different semantics. In spite of that the verification conditions have a similar nature to ours and give rise to similar parametric set constraints. Our work goes further in that we use such parametric constraints in a sufficient correctness test, and also for type inference, while the simplification techniques of (Boye, 1996) are rather limited in handling parameters.

The problem of polymorphic directional type checking is also addressed in (Rychlikowski & Truderung, 2000) and more recently in (Rychlikowski & Truderung, 2001). This work presents a formal system, where directional well-typing of a logic program for given type specification is defined in terms of proofs constructed from given axioms and typing rules. This is different from our approach where the well typing algorithms are derived from the semantic concept of program correctness and types are understood as families of sets, specified by means of PED-grammars. Thus it seems impossible to compare our type checking algorithms with those discussed in (Rychlikowski & Truderung, 2000).

Nevertheless, the semantics of types as sets is also provided in (Rychlikowski & Truderung, 2000). It is done by a fixpoint construction, which for a given alphabet of typed function symbols associates each used type with a subset of the Herbrand universe. In this, rather indirect, way a similar effect is obtained as by our direct specification of types by means of PED-grammars. However, the class of the sets which can be constructed in that way is not precisely characterized. Syntactic restrictions on the way of defining signatures seem to make it somewhat restricted. For example, it is impossible to have nonempty intersection of instances of different polymorphic types, e.g. [] cannot be used for representing both the empty list and the empty tree. This is a substantial restriction, e.g. one cannot define a type of even length lists.

The soundness theorem of (Rychlikowski & Truderung, 2000) relates the directional types of well-typed programs to their declarative semantics, while the types discussed here are related to the call-success semantics. Failure of our type checking algorithm locates potential errors in a fragment of a clause, while a proof failure of (Rychlikowski & Truderung, 2000) seems to indicate a whole clause. (At least this issue is not discussed in that paper.) Handling of constraints is not discussed in their work, its main objective is representing different directional types of a predicate by one *main* type.

### *7.2 Conclusions*

We extended the concept of partial correctness for a logic program wrt to a directional type to a concept of partial correctness of a CLP program wrt to a parametric directional specification. We formulated sufficient conditions for the correctness and we encoded them as set constraints. In this way we gave a semantic-based view of parametric polymorphism in constraint logic programs.

We extended the notion of discriminative term grammar to the notion of parametric extended discriminative term grammar (PED grammar). We argued that directional types specified by such grammars are quite useful. On one hand, they make it possible to describe simple approximations of program semantics, easy to provide and to understand by the user. On the other hand, they allow automatic check of the sufficient conditions mentioned above. Using these conditions one can also automatically infer parametric directional types from a parametric entry declaration.

Our type inference techniques extend to CLP and to the parametric types the techniques of (Gallagher & de Waal, 1994) corrected by Mildner (1999); they are based on abstract interpretation of logic programs. It seems possible to extend instead some of the set constraint solving techniques. This may be a topic of future work including also a comparison of both extensions.

We developed a prototype tool implementing the proposed algorithms, which can be obtained from the third author. The theoretical result of (Charatonik & Podelski, 1998) shows that the problem of checking discriminative directional types is not tractable, even in the parameterless case. The complexity of our type checking algorithm is exponential w.r.t. the maximal number of occurrences of a variable in a clause. However our tool turns out to be sufficiently efficient for practical purposes.

Our tool supports a compile-time technique for error location based on checking directional parametric types. Clearly, the class of errors that can be located is restricted to type errors. The check locates those clause prefixes, which cause the type errors. Our approach does not impose any type discipline on the program. It does not require providing all type declarations in advance and often only a few declarations are sufficient to locate an error. The process of specifying declarations is supported by the possibility of inspecting and adopting the inferred types.

### Acknowledgments

We want to thank the anonymous referees for insightful comments and suggestions which resulted in substantial improvements of this paper. We acknowledge the contribution of Marco Comini who was largely involved in the development of an early version of the diagnosis tool. We also thank our colleagues from the DiSCiPl Project in which this research has been initiated. In particular we are grateful to Pierre Deransart and Manuel Hermenegildo for fruitful discussions. This work was partly funded by The Swedish Research Council grant 1999-109.